\documentclass[journal]{IEEEtran}
\usepackage{diagbox}
\usepackage[inline]{enumitem}
\usepackage{xcolor}
\usepackage{amsmath,amsfonts}
\usepackage{algorithm}
\usepackage[group-separator={,},group-minimum-digits=4]{siunitx}
\usepackage{threeparttable}
\usepackage{array}
\usepackage[caption=false,font=normalsize,labelfont=sf,textfont=sf]{subfig}
\usepackage{textcomp}
\usepackage{stfloats}
\usepackage{url}
\usepackage{verbatim}
\usepackage{graphicx}
\usepackage{algpseudocode}
\usepackage{cite}
\usepackage{balance}
\usepackage{mathrsfs}
\usepackage[colorlinks=true,
            linkcolor=blue,
            anchorcolor=blue,
            citecolor=blue]{hyperref}

\begin{document}

\title{
{\fontsize{20.5pt}{\baselineskip}\selectfont Anti-Jamming Precoding Against Disco Intelligent Reflecting Surfaces Based Fully-Passive Jamming Attacks}
} 
\author{
        Huan~Huang,~\textit{Member, IEEE},
        Lipeng~Dai,
        Hongliang~Zhang,~\textit{Member, IEEE},
        Zhongxing~Tian,
        Yi~Cai,~\textit{Senior~Member,~IEEE},
        Chongfu~Zhang,~\textit{Senior~Member,~IEEE},\\
        A.~Lee~Swindlehurst,~\textit{Fellow,~IEEE},
        and~Zhu~Han,~\textit{Fellow,~IEEE}
\thanks{
A portion of this work was published in~\cite{MyGC23}.

This work was supported by the National Key R\&D Program of China (2022YFB2903000) 
and the National Natural Science Foundation of China (62275185, 62250710164), and partially supported by the U.S. National
Science Foundation (CNS--2107216, CNS--2128368, CMMI--2222810, ECCS--2030029, CNS--2107182), 
US Department of Transportation, Toyota, and Amazon (\textit{Corresponding author: Yi~Cai}).

H.~Huang, Z.~Tian, and Y.~Cai are with the School of Electronic and Information Engineering, 
Soochow University, Suzhou 215006, China 
(e-mail: hhuang1799@gmail.com, zxtian@ieee.org, yicai@ieee.org).

L.~Dai and C.~Zhang are with the School of Information and Communication Engineering, 
University of Electronic Science and Technology of China, Chengdu 611731, China 
(e-mail: dlp1022@163.com, cfzhang@uestc.edu.cn).

H.~Zhang is with the School of Electronics, Peking University, Beijing 100871, China 
(e-mail: hongliang.zhang92@gmail.com).

A.~L.~Swindlehurst is with the Center for Pervasive Communications and Computing, 
University of California, Irvine, CA 92697, USA (e-mail: swindle@uci.edu).

Z.~Han is with the University of Houston, Houston, TX 77004 USA 
(e-mail: hanzhu22@gmail.com).}
}
\maketitle

\begin{abstract}
Emerging intelligent reflecting surfaces (IRSs) significantly 
improve system performance, but also pose a huge risk for physical 
layer security. Existing works have illustrated that a disco IRS (DIRS), 
i.e., an illegitimate IRS with random time-varying reflection properties 
(like a ``disco ball"), can be employed by an attacker to actively 
age the channels of legitimate users (LUs).
Such active channel aging (ACA) generated by the 
DIRS  
can be employed to jam multi-user multiple-input single-output (MU-MISO) systems 
without relying on either jamming power or LU channel state information (CSI). To address 
the significant threats posed by DIRS-based fully-passive jammers (FPJs), an 
anti-jamming precoder is proposed that requires only the statistical characteristics of 
the DIRS-based ACA channels instead of their CSI. The statistical characteristics of 
DIRS-jammed channels are first derived, and then the anti-jamming precoder is derived 
based on the statistical characteristics. Furthermore, we prove that the anti-jamming precoder can achieve the maximum signal-to-jamming-plus-noise ratio (SJNR). To acquire the ACA statistics without changing the system architecture or cooperating with the illegitimate DIRS, we design a data frame structure that the legitimate access point (AP) can use to estimate the statistical characteristics. During the designed data frame, the LUs only need to feed back their received power to the legitimate AP when they detect jamming attacks.
Numerical results are also presented to evaluate the effectiveness of 
the proposed anti-jamming precoder against the DIRS-based FPJs and the 
feasibility of the designed data frame used by the legitimate AP to estimate the statistical characteristics.
\end{abstract}


\section{Introduction}\label{Intro}
Due to the broadcast and superposition properties of wireless channels, wireless communications are vulnerable to malicious attacks such as eavesdropping and jamming~\cite{PLSsur1,DoSsur1,AntiJammingSurv,addJE}. To protect legitimate users (LUs) from eavesdropping, cryptographic techniques are used to prevent eavesdroppers from intercepting transmitted signals~\cite{PLSsur1,DoSsur1}. Cryptographic techniques for secure communications rely on the computational difficulty of the underlying mathematical process required to break the codes. Therefore, the eavesdroppers can only effectively receive the transmit signals if they have extensive computational capabilities~\cite{crypbook}.

On the other hand, jamming attacks (also referred to as DoS-type attacks) can be launched by an active jammer (AJ) that imposes intentional interference on the communication between the legitimate access point (AP) and its LUs~\cite{AntiJammingSurv}. In practice, physical-layer AJs can generally be classified into constant AJs, intermittent AJs, reactive AJs, and adaptive AJs~\cite{DoSsur1}. A constant AJ continuously broadcasts jamming signals, such as  modulated Gaussian waveforms or pseudorandom noise, over an open wireless channel to prevent LUs from communicating with the legitimate AP.
However, constant AJs are energy-inefficient because they constantly consume power, and thus energy constraints are an inherent drawback for AJs~\cite{DoSJamm}. To overcome this drawback, intermittent AJs~\cite{intermittentAJ}, reactive AJs~\cite{reactiveAJ}, and adaptive AJs~\cite{adaptiveAJ} have been investigated. The basic idea of these AJs is to reduce the duration of the jamming transmission in order to reduce the consumption of power.
However, all types of active jamming require a certain amount of jamming power to effectively attack the LUs. Given the inherent energy disadvantage of AJs, can jamming attacks be launched without jamming power?

Recently, intelligent reflecting surfaces (IRSs) have been considered to be a promising technology for future 6G systems, and can be used to reflect electromagnetic waves in a controlled manner~\cite{IRSsur1,HuangRIS}. 
Specifically, an IRS is an ultra-thin surface equipped with multiple sub-wavelength reflecting elements whose electromagnetic responses (i.e., amplitudes and phase shifts) can be controlled, for instance, by simple programmable PIN or varactor diodes~\cite{IRSsur2}. 
Previous works have mainly focused on the use of legitimate IRSs in order to improve performance metrics such as spectrum efficiency (SE)~\cite{MySE,AORIS,CC1}, energy efficiency (EE)~\cite{MyEE}, 
cell coverage~\cite{PostDrRenQ}, or  physical-layer security~\cite{AddPLS1,AddPLS2},
assuming the legitimate AP knows the the IRS-related channel state information (CSI), and can control their phase response. 
The authors of~\cite{CC1,PostDrRenQ} significantly enhanced the coverage and spectrum efficiency via the practical design of reconfigurable surfaces.

However, the emergence of IRSs also poses serious potential threats to wireless networks. Some works have pointed out that illegitimate IRSs can have a significant impact on wireless networks 
because the illegitimate IRSs~\cite{IIRSSur,IIRSSuradd,IIRSSuradd1,IIRSSuradd2} are difficult to detect due to their passive nature. 
For example, the authors in~\cite{PassJamSU} have reported an adversarial IRS-based passive jammer that essentially consumes no jamming power and 
that can destructively add the signal reflected from it with the direct path signal to minimize the received power at the LU, i.e., the signal-to-noise ratio (SNR). Although this passive jammer can launch jamming attacks without jamming power, the CSI of all wireless channels must be known at the unauthorized  IRS. 
Due to the passive nature of IRSs, the CSI of IRS-aided channels is estimated jointly with that of the legitimate AP and LUs. 
Specifically, by exploiting the channel reciprocity of time division duplex (TDD) channels, the LUs instead of the legitimate IRS send pilot signals to the AP, and the AP then estimates the IRS-aided channels using methods such as the least squares (LS) algorithm~\cite{DaipartI}. 
If the illegitimate IRS aims to acquire LU CSI, it must train to learn CSI jointly with the legitimate AP and LUs. 
As a result, the assumption that the illegitimate IRS knows the CSI~\cite{PassJamSU} is unrealistic for practical wireless networks.

Considering the difficulty of illegitimate IRSs to acquire CSI, can jamming attacks be launched without either jamming power or LU CSI? 
An interesting fully-passive jammer (FPJ)~\cite{DIRSVT,DIRSTWC} has been proposed to launch jamming attacks on LUs with neither LU CSI nor jamming power, where an illegitimate IRS with random phase shifts, referred to as a ``disco" IRS (DIRS), is used to actively age the LUs' channels. This causes serious active channel aging (ACA) interference, which is a type of inter-user interference (IUI).
Specifically, the DIRS controller in~\cite{DIRSVT} randomly generates a reflecting vector once during the \emph{reverse pilot transmission (RPT)} phase. 
Then, during the subsequent \emph{data transmission (DT)} phase, the DIRS controller randomly generates another reflecting vector. 
In~\cite{DIRSTWC}, the authors further illustrated that the DIRS-based ACA interference can also be introduced by turning off the illegitimate IRS during the \emph{RPT} phase and then randomly generating reflecting vectors multiple times during the \emph{DT} phase. 
Such a temporal DIRS-based FPJ must know when the \emph{RPT} phase ends and the \emph{DT} phase begins, which requires some synchronization. 
These DIRS-based FPJs impose significant risks to PLS since they are difficult to detect due to their passive nature.
It is worth noting that the adversarial IRSs studied in~\cite{DIRSVT,DIRSTWC} 
address a problem that is different from that in~\cite{AddR3}, where an adversarial IRS is employed to launch pilot contamination attacks 
to improve the eavesdropping capability of Eve. 
Moreover, these different RIS-based attack strategies are summarized in Table~\ref{tab1add}.

\begin{table*}
    \footnotesize
    \centering
    \caption{Comparison of Different RIS-Based Attack Strategies}
    \label{tab1add}
    \begin{tabular}{ ||c||c|c|c| }
    \hline
    Reference                    &\cite{PassJamSU}               &\cite{IIRSSuradd1,IIRSSuradd2,DIRSVT,DIRSTWC}          &\cite{AddR3}  \\
    \hline
    Attack type                  &Jamming                        &Jamming                                                &Eavesdropping\\
    \hline
    Transmit energy              &Not required                   &Not required                                           &Not required\\
    \hline
    Channel knowledge            &Required                       &Not required                                           &Not required\\
    \hline
     Mechanism                   &Optimize RIS reflecting        &Generate time-varying                                  &Reflect pilots by RIS to\\ 
     {}                          &vector to minimize SNR         &RIS reflecting vectors                                 &enhance  eavesdropping\\
    \hline
    \end{tabular}
\end{table*}


In conventional wireless networks,
classical anti-jamming approaches~\cite{AntiJammingSurv} such as spread spectrum and frequency-hopping techniques have been widely used to suppress jamming attacks.
Spread spectrum refers to spreading the signal energy over a wider range of frequencies than the minimum required for transmission. In addition, frequency hopping is a technique used in spread spectrum communications in which the carrier frequency is rapidly changed in a pattern known to both the transmitter and receiver.
However, classical anti-jamming approaches such as these can not be used against an FPJ since the source of the jamming attacks launched by the FPJs comes from the legitimate AP transmit signals themselves, which have the same characteristics (e.g., carrier frequency) as the transmit signals.

In addition, the ACA interference from the DIRS-based FPJs cannot be mitigated using multi-input multi-output (MIMO) interference cancellation~\cite{JammResilientCommunConf,JammResilientCommun}. MIMO interference cancellation is effective for DIRS-based ACA interference only if the channel information of both the LU and DIRS-jammed channels is known by the legitimate AP~\cite{JammResilientCommunConf,JammResilientCommun}. However, the DIRS phase shifts and amplitudes are randomly generated~\cite{DIRSVT,DIRSTWC}. It has been shown in~\cite{DIRSTWC} that a DIRS-based FPJ using only one-bit quantized phase shifts can achieve the desired jamming effect as long as the number of DIRS elements is sufficiently large. The key advantage of these approaches is that there is no effective anti-jamming approach available to counteract these destructive jamming attacks imposed by DIRS-based FPJs~\cite{DIRSVT,DIRSTWC}.

To respond to the significant risks posed by illegitimate IRSs, an anti-jamming precoder is proposed in this paper for attacks launched by DIRS-based FPJs, which only requires the statistical characteristics of the DIRS-based ACA channels instead of their instantaneous CSI.
The main contributions are summarized as follows:

\begin{itemize}
\item A practical IRS model is considered, where the phase shifts of the DIRS reflecting elements are discrete and the amplitudes are a function of their corresponding phase shifts. Based on this practical IRS model, we describe a persistent DIRS-based FPJ that initiates jamming attacks through DIRS-based ACA interference and requires no additional jamming power or knowledge of the LU CSI, where the DIRS phase shifts are randomly generated once during the \emph{RPT} phase and then randomly generated multiple times during the \emph{DT} phase. Compared to~\cite{MyGC23,DIRSTWC}, the persistent DIRS-based FPJ mode$\footnote{The DIRS controller for the case considered in~\cite{DIRSTWC,MyGC23} generates a zero reflecting vector (i.e., the wireless signals are perfectly absorbed by the DIRS) during the \emph{RPT} phase. Therefore, we refer to the approach studied here as persistent DIRS-based FPJ.}$ is more harmful because it is not necessary to be synchronized with the training process of the legitimate system. Therefore, the developed anti-jamming precoder is more comprehensive. Moreover, the persistent DIRS-based FPJ based on the practical IRS model has different properties compared to~\cite{DIRSVT,DIRSTWC}, for instance, the distribution of the random DIRS phase shifts affects the jamming impact of the persistent DIRS-based FPJ.
\item To address the serious threats posed by DIRS-based FPJs, we develop an anti-jamming precoder that requires only the statistical characteristics of the DIRS-based ACA channels and avoids requiring their instantaneous CSI, which is impractical to obtain. First, we derive the statistical characteristics of the DIRS-based ACA channels for both the persistent DIRS-based FPJ and the temporal DIRS-based FPJ~\cite{DIRSTWC,MyGC23}. Based on the derived characteristics, we explain the difference between the jamming impact of the two methods. Second, we develop an anti-jamming precoder based only on the statistical characteristics, and prove that this precoder can achieve the maximum signal-to-jamming-plus-noise ratio (SJNR). The proposed anti-jamming strategy works for both the persistent DIRS-based FPJ and the related DISCO approaches in~\cite{DIRSVT,DIRSTWC}.
\item For practical applications, it is necessary for the legitimate AP to acquire the ACA statistical characteristics without changing the system architecture or cooperating with the illegitimate DIRS. To this end, we design a data frame structure that the legitimate AP can use to estimate the statistical characteristics. Specifically, the LUs only need to feed back their received power to the legitimate AP when they detect that jamming is present. This requires little overhead because the received power values of the LUs are scalars and only a few feedback transmissions are sufficient to effectively estimate the statistical characteristics.
\end{itemize}

\begin{figure}[!t]
    \centering
    \includegraphics[scale=0.6]{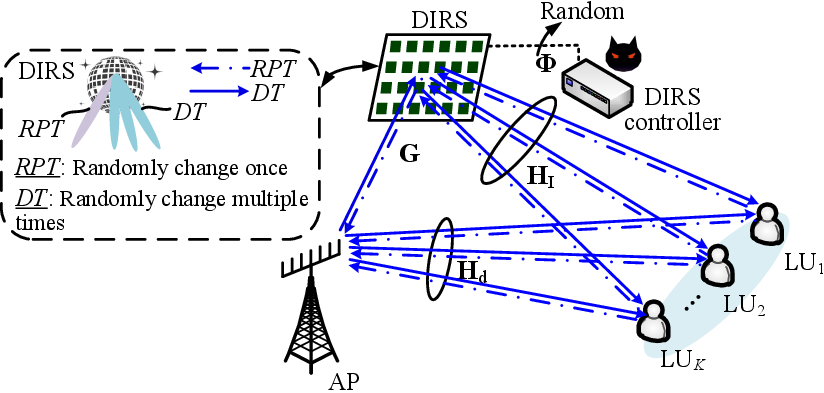}
    \caption{The downlink of a legitimate MU-MISO system jammed by a persistent disco intelligent reflecting surface based fully-passive jammer (DIRS-based FPJ), where the DIRS reflecting vectors during the \emph{reverse pilot transmission (RPT)} and \emph{data transmission (DT)} phases are randomly and independently generated by the DIRS controller.}
    \label{fig1}
\end{figure}

The rest of this paper is organized as follows. 
In Section~\ref{Princ}, we present the downlink of an MU-MISO system jammed by a persistent DIRS-based FPJ and define the SJNR optimization metric to quantify the anti-jamming effect. In addition, some useful results on matrix analysis and random variables are reviewed. In Section~\ref{AntiJamming}, the statistical characteristics of the DIRS-based ACA channels are derived for the persistent DIRS-based FPJ and the earlier approach in~\cite{DIRSTWC,MyGC23}. Then, an anti-jamming precoder is designed based on the derived statistical characteristics, and we prove that this precoder can achieve the maximum SJNR. In Section~\ref{AntiFrame}, we develop a data frame structure the legitimate AP can use to estimate the statistical characteristics and we explain the mechanism by which it works. 
Finally, conclusions are given in Section~\ref{ResDis}, 
where the difference between the persistent DIRS-based FPJ 
and the version in~\cite{DIRSTWC,MyGC23} is compared and discussed in detail.

\emph{Notation:} We employ bold capital letters for a matrix, e.g., ${\bf{W}} $, lowercase bold letters for a vector, e.g., $\boldsymbol{w} $, and italic letters for a scalar, e.g., $K$. The superscripts $(\cdot)^{-1}$, $(\cdot)^{T}$, and $(\cdot)^{H}$ represent the inversion, the transpose, and the Hermitian transpose, respectively, and the symbols $\|\cdot\|$ and $|\cdot|$  represent the Frobenius norm and the absolute value, respectively. 

\section{System Description}\label{Princ}
In Section~\ref{DIRSFPJ}, based on a practical IRS model, we illustrate the downlink of an MU-MISO system jammed by a persistent DIRS-based FPJ. Then, we define the SJNR optimization metric to quantify the system performance under this persistent DIRS-based FPJ. In Section~\ref{ChanMod}, the wireless channels involved 
are modelled based on near-field and far-field models, as appropriate. In Section~\ref{RandomVar}, some important results on matrix analysis and random variables are reviewed, which will be useful for the anti-jamming precoding derived in Section~\ref{AntiJamming}.

\subsection{MU-MISO Systems Jammed by DIRS-based Fully-Passive Jammers}\label{DIRSFPJ}
Fig.~\ref{fig1} schematically shows the general downlink model of an MU-MISO system that is jammed by the persistent DIRS-based FPJ. A legitimate AP uses $N_{\rm A}$ antennas to communicate with $K$ single-antenna legitimate users denoted by LU$_1$, $\cdots$, LU$_K$. Meanwhile, a DIRS with $N_{\rm D}$ reflecting elements is employed to launch fully-passive jamming attacks on the LUs. In many existing IRS-enhanced systems, it is assumed that the IRSs are placed close to the users to maximize a certain performance metric~\cite{MyEE,MySE,AORIS,IRSdeployment,PostDrRenQ}. However, the assumption that the illegitimate DIRS has no information about the LUs, such as the LUs' locations, is more realistic in the jamming scenario~\cite{DIRSVT,DIRSTWC,MyGC23}. Therefore, the DIRS is assumed to be deployed close to the legitimate AP to maximize the impact of the DIRS~\cite{IRSdeployment}.

\begin{figure}[!t]
    \centering
    \captionsetup[subfloat]{labelsep=none,format=plain,labelformat=empty}
    \setlength{\abovecaptionskip}{0cm}
      \subfloat[]{
          \includegraphics[width=0.6\linewidth]{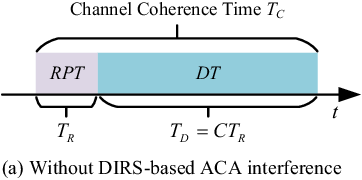}}\vspace{-0.4cm}
          \label{fig2a}
      \subfloat[]{
          \includegraphics[width=0.6\linewidth]{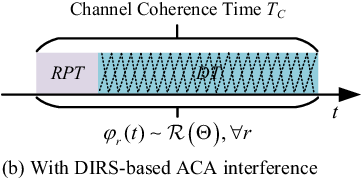}}
          \label{fig2b}
      \caption{Schematic diagram of the frames of an MU-MISO system (a) without jamming attacks and (b) with jamming attacks launched by the persistent DIRS-based FPJ.}
\label{fig2}
\end{figure}

\underline{\textit{Disco Intelligent Reflecting Surfaces:}}
In general, in the downlink of an MU-MISO system, the legitimate AP jointly trains the CSI with the LUs during the \emph{RPT} phase, 
and the legitimate AP then designs a precoder that is used to transmit signals to the LUs during the \emph{DT} phase. 
As shown in Fig.~\ref{fig2}(a), the channel coherence time consists of two phases, i.e., the \emph{RPT} phase and the \emph{DT} phase, and the \emph{DT} phase generally lasts much longer than the \emph{RPT} phase.
We assume that the length of the \emph{RPT} phase is $T_{R}$ and that of the \emph{DT} phase is $T_{D} = CT_R$ for some integer $C > 1$.
Since existing MU-MISO systems assume that the wireless channels remain unchanged during the channel coherence time, 
the designed precoder, such as the widely-used zero-forcing (ZF) precoder~\cite{ZFBF}, can achieve good performance.
However, the works in~\cite{DIRSVT,DIRSTWC} have shown that an attacker can exploit the ability of the IRS to controllably change the wireless channels to launch jamming attacks with neither jamming power nor LU CSI.



Unlike the temporal DIRS-based FPJ presented in~\cite{MyGC23,DIRSTWC}, the fully-passive jamming attacks can also be launched by randomly changing the DIRS reflection coefficients once in the \emph{RPT} phase and multiple times in the \emph{DT} phase, as depicted in Fig.~\ref{fig2}(b).
When the DIRS controller detects that the MU-MISO system begins to communicate, it controls the DIRS reflection properties to change randomly.
The period during which the DIRS reflection coefficients are changing is about the same as the length of the \emph{RPT} phase, i.e., $T_{R}$. 
Therefore, during each \emph{DT} phase, the first random change in the DIRS reflection coefficients starts at the same time as the \emph{DT} phase.
It is worth noting that the basic idea of DIRSs is to rapidly age the CSI during the channel coherence time instead of preventing the legitimate AP from obtaining it. As a result, signals transmitted to LUs during the \emph{DT} phase will be jammed by the persistent DIRS-based FPJ.

Compared to the temporal DIRS-based FPJ in~\cite{MyGC23,DIRSTWC}, the proposed persistent DIRS-based FPJ does not require synchronization. In practice, it is difficult for an illegitimate DIRS to access information about the training synchronization of the legitimate MU-MISO system.
Moreover, the difference between the jamming impact of the temporal DIRS-based FPJ in~\cite{MyGC23,DIRSTWC} and that of the persistent DIRS-based FPJ shown in Fig.~\ref{fig2} (b) is quantified based on the theoretical derivation in Section~\ref{AntiPrecoder}.
The ACA from both the persistent DIRS-based FPJ and the temporal DIRS-based FPJ~\cite{MyGC23,DIRSTWC} is different from the channel aging (CA) in a traditional MU-MISO system, which is caused by time variations in the RF propagation and delays in computation between the time the channels are learned at the legitimate AP and when they are used for precoding~\cite{ChanAge}. This type of CA can not be actively introduced and controlled.

In practice, an IRS is an ultra-thin surface equipped with multiple sub-wavelength reflecting elements whose phase shifts and amplitudes can be controlled by simple programmable PIN or varactor diodes~\cite{IRSsur2}. We will assume the use of PIN diodes, whose ON/OFF behavior only allows for the creation of discrete phase shifts. In addition, due to the properties of the IRS, the amplitudes of the reflecting elements are a function of their corresponding phase shifts~\cite{IRSsur1}.
In particular, assume that the DIRS has $b$-bit quantized phase shifts whose 
values can be chosen from the set $\Theta = \left\{ {{\theta _1}, \cdots ,{\theta _{{2^b}}}} \right\}$, then the time-varying DIRS reflecting vector is given by
\begin{equation}
{\boldsymbol{\varphi}}(t) = \left[ {{\alpha}_1}{\!(t)}{{e^{j{\varphi _1}(t)}}, \cdots ,{{\alpha}_{N\!_{\rm D}}}{\!(t)}{e^{j{\varphi\!_{{N\!_{\rm{D}}}}}\!(t)}}} \right],
\label{DIRSPhase}
\end{equation}
where the phase shift of the $r$-th DIRS reflecting element is randomly selected from the possible phase shift set $\Theta$, i.e., ${\varphi _r}(t) \sim {\cal R}\!\left(\Theta\right)$. The probability of the phase shift ${\varphi _r}(t)$ taking the $i$-th value of $\Theta$ is represented by $P_{r,i}$, i.e., $P_{r,i} = {\mathbb{P}}\!\left( {\varphi _r}(t) = {\theta _{i}} \right) = P_{i}$ and $i = 1,\cdots, 2^b$.
Furthermore, the corresponding gain value ${{\alpha}_r}{(t)}$ is a function of ${\varphi _r}(t)$ and denoted by ${{\alpha}_r}{(t)} = {\cal F}\!\left({\varphi _r}(t)\right)$. We further denote all possible gain values by $\Omega = {\cal F}\!\left({\Theta}\right) = \left\{{\mu _1},\cdots,{\mu_{2^b}} \right\}$.

The work in~\cite{DIRSTWC} has shown that the jamming impact of a temporal DIRS-based FPJ does not depend on how the discrete random DIRS phase shifts are distributed when the element gains of the DIRS are assumed to be the same, i.e., the constant-amplitude assumption. However, we will show in the following section that this conclusion no longer holds when the gains of the DIRS reflecting elements are different.

\underline{\textit{Reverse Pilot Transmission And Linear Precoder Design:}}
As mentioned above, the LUs' CSI is obtained in the \emph{RPT} phase through joint training with the LUs in order for the AP to design a  precoder that is used to transmit LU signals during the \emph{DT} phase~\cite{PilotConta}. Specifically, the pilot signal transmitted by LU$_k$ is denoted by $s_{{\!R\!P\!T},k}$. During the \emph{RPT} phase, the received pilot vector ${\boldsymbol{y}_{{\!R\!P\!T},k}} \in \mathbb{C}^{{N_{\rm A}}\times 1}$ at the AP  is
\begin{alignat}{1}
\nonumber
{\!\boldsymbol{y}_{{\!R\!P\!T},k}} &= \sqrt {{p_{{\!R\!P\!T},k}}}{\boldsymbol{h}_{{\!R\!P\!T},k}}{s_{{\!R\!P\!T},k}} + {\boldsymbol{n}_{{\!R\!P\!T},k}} \\
& = \sqrt {{p_{{\!R\!P\!T},k}}} \! \left(  { {\boldsymbol{h}_{{\rm{I}},k} {{\bf \Phi}_{\!R\!P\!T}}{{\bf{G}}}} + \boldsymbol{h}_{{\rm{d}},k} } \right) \!{s_{{\!R\!P\!T},k}} + {\boldsymbol{n}_{{\!R\!P\!T},k}},
\label{PTeq}
\end{alignat}
where ${p_{{\!R\!P\!T},k}}$ is the transmit power of 
$s_{\!R\!P\!T,k}$, $\boldsymbol{h}_{{\rm{I}},k} \in \mathbb{C}^{ 1 \times {N_{\rm D}}}$ 
denotes the channel between the DIRS and LU$_k$, 
${\bf{G}} \in \mathbb{C}^{ {N_{\rm D}} \times {N_{\rm A}}}$ 
denotes the channel between the DIRS and the AP, 
and $\boldsymbol{h}_{{\rm{d}},k} \in \mathbb{C}^{ 1 \times {N_{\rm A}}}$ 
denotes the direct channel between the AP and LU$_k$. 
In addition, ${\boldsymbol{n}_{{\!R\!P\!T},k}} = {\left[ {{n_{k,1}}, \cdots, {n_{k,{N_{\rm A}}}}} \right]^T}$ 
denotes additive white Gaussian noise (AWGN) at the AP that consists of independent and identically distributed (i.i.d.) elements with zero mean and variance $\sigma^2$, i.e., ${{n}_{k,i}} \sim \mathcal{CN}\left(0,\sigma^2\right)$, $i = 1, \cdots ,{N_{\rm A}}$. For ease of presentation, we denote the DIRS-jammed channel between the AP and LU$_k$ by $\boldsymbol{h}_{{\rm{D}},k}^{R\!P\!T} = {\boldsymbol{h}_{{\rm{I}},k} {{\bf \Phi}_{\!R\!P\!T}} {{\bf{G}}}}$, where ${{\bf \Phi}_{\!R\!P\!T}} = {{\rm{diag}}\!\left( {\boldsymbol{\varphi}}(t_{\!R\!P\!T}) \right)}$ represents the DIRS passive beamforming during the \emph{RPT} phase. The DIRS passive beamforming $\boldsymbol{h}_{{\rm{D}},k}^{R\!P\!T}$ remains unchanged during the \emph{RPT} phase in a channel coherence interval, but changes during the \emph{RPT} phases belonging to different channel coherence intervals.
The overall DIRS-LU channel ${{\bf H}}_{\rm I}$ and the overall direct channel ${{\bf H}}_{\rm d}$ are represented by ${\bf H}_{{\rm I}}^H = \left[\boldsymbol{h}_{{\rm{I}},1}^H,\cdots,\boldsymbol{h}_{{\rm{I}},K}^H  \right]$ and ${\bf H}_{{\rm d}}^H = \left[\boldsymbol{h}_{{\rm{d}},1}^H,\cdots,\boldsymbol{h}_{{\rm{d}},K}^H  \right]$, respectively.

In IRS-aided systems, the AP controls the legitimate IRS, and thus the IRS-aided and direct channels can be estimated~\cite{DaipartI}.
However, in the jamming scenario investigated here and in the literature~\cite{DIRSVT,DIRSTWC}, the DIRSs are controlled by a malicious adversary. 
Therefore, the AP can only estimate the combined channel ${\boldsymbol{h}_{{\!R\!P\!T},k}} \in \mathbb{C}^{{N_{\rm A}}\times 1}$ according to the received pilot vector ${\boldsymbol{y}_{{\!R\!P\!T},k}}$. 
Similarly, the combined channels between the AP and all LUs can be obtained in the $RPT$ phase. 
The overall combined channel ${\bf H}_{\!R\!P\!T} \in \mathbb{C}^{ {N_{\rm A}} \times {K}}$ is written as ${\bf H}_{\!R\!P\!T}  = \left[{\boldsymbol{h}_{{\!R\!P\!T},1}}, \cdots, {\boldsymbol{h}_{{\!R\!P\!T},K}}  \right]$, 
and knowledge of ${\bf H}_{\!R\!P\!T}$ can be obtained by the legitimate AP because the random DIRS reflection coefficients are constant during the \emph{RPT} phase. 
Herein, we assume that the CSI of ${\bf H}_{\!R\!P\!T}$ can be obtained by the AP during the $RPT$ phase~\cite{PerfectIRSCSI}, 
as imperfect CSI is not a primary concern in the jamming scenario, and its impact has also been thoroughly studied~\cite{LDetector1,RefSLNR,RefSLNRadd}.

According to the obtained ${\bf H}_{\!R\!P\!T}$, the AP then designs a precoder that is used to transmit signals to the LUs during the $DT$ phase. In general, the aim of an MU-MISO system is to maximize desired signals and minimize IUI. A widely-used linear precoder that can achieve zero IUI is the zero-forcing (ZF) algorithm~\cite{ZFBF}. Specifically, based on ${\bf H}_{\!R\!P\!T}$, the ZF precoder used at the AP can be computed by
\begin{equation}
{{\bf{W}}\!_{\rm{ZF}}} \!=\! {{{{\bf{H}}\!_{{R\!P\!T}}}{{\!\left( {{\bf{H}}\!_{{R\!P\!T}}^H{{\bf{H}}\!_{{R\!P\!T}}}} \right)}^{ - 1}}{{\bf{P}}^{ \frac{1}{2}}}}} \!=\! \left[{{\boldsymbol{w}}\!_{{\rm{ZF}},1}}, \cdots, {{\boldsymbol{w}}\!_{{\rm{ZF}},K}}\right],
\label{ZF}
\end{equation}
where ${\bf{P}} = {\rm{diag}}\!\left(p_1,\cdots, p_K\right)$ is the power allocation matrix,  $\|{{\boldsymbol{w}}\!_{{\rm{ZF}},k}}\| = \sqrt{p_{k}} $, and $p_k$ denotes the transmit power allocated to ${\rm{LU}}_k$. The total transmit power $P_0$ used by the legitimate AP to transmit signals satisfies  $\sum\nolimits_{k = 1}^K {{p_k}}  \le {P_0}$. For simplicity, we further assume that $p_{k} = \frac{P_{0}}{K}, \forall k$.

\underline{\textit{Data Transmission And Active Channel Aging Interference:}} Once the precoder has been computed, the legitimate AP uses this precoder to transmit signals to the LUs during the \emph{DT} phase. Assuming that the transmit signal for LU$_k$ satisfies ${\mathbb{E}}\!\left[ \left| s_{{\!D\!T},k} \right|^2 \right] = 1$, the signal received at ${\rm{LU}}_k$ during the \emph{DT} phase is given by
\begin{alignat}{1}
\nonumber
{y_{\!D\!T,k}} &= \boldsymbol{h}\!_{D\!T,k}^H\!\sum_{u = 1}^K {{\boldsymbol{w}_{{\rm{\!Z\!F}},u}}{s_{\!D\!T,u}}}  + {n_k}\\
&=    \left({{{{\bf{G}}^H} {\!{\bf{\Phi}}^H_{\!D\!T}} \boldsymbol{h}_{{\rm{I}},k}^H} + \boldsymbol{h}_{{\rm{d}},k}^H} \right)\!{\sum_{u =1}^K}{{\boldsymbol{w}_{{\rm{\!ZF}},u}}{s_{\!D\!T,u}}} + {n_k} ,
\label{RecSig}
\end{alignat}
where ${{\bf \Phi}_{\!D\!T}} = {{\rm{diag}}\!\left( {\boldsymbol{\varphi}}(t_{\!D\!T}) \right)}$ represents the DIRS passive beamforming during the $DT$ phase and the AWGN ${n_k}$ received at LU$_k$ is also assumed to have zero mean and variance  $\sigma^2$, i.e., ${{n}_{k}} \sim \mathcal{CN}\left(0,\sigma^2\right)$. Furthermore, we denote the DIRS-jammed channel between the AP and LU$_k$ during the \emph{DT} phase by $\boldsymbol{h}_{{\rm{D}},k}^{D\!T} = {\boldsymbol{h}_{{\rm{I}},k} {{\bf \Phi}_{\!D\!T}} {{\bf{G}}}}$.

Due to the change in the DIRS reflecting vector between the \emph{RPT} phase and the \emph{DT} phase, 
there is a difference between the obtained overall combined 
channel ${\bf H}_{\!R\!P\!T}$ in the \emph{RPT} phase and the actual overall combined channel ${\bf H}_{\!D\!T}$ during the \emph{DT} phase. Mathematically, the DIRS-based ACA channel ${{\bf{H}}_{\rm{\!A\!C\!A}}}$ is expressed by
\begin{equation}
{{\bf{H}}_{\rm{\!A\!C\!A}}} =  {{\bf{H}}_{\!D\!T}} - {{\bf{H}}_{\!R\!P\!T}}  =\left[{{\boldsymbol{h}}_{{\rm{\!A\!C\!A}},1}}, \cdots, {{\boldsymbol{h}}_{{\rm{\!A\!C\!A}},K}}\right],
\label{ChannelACA}
\end{equation}
where ${{\bf{H}}_{\!D\!T}} = \left[{\boldsymbol{h}_{\!D\!T,1}},\cdots,{\boldsymbol{h}_{\!D\!T,K}}\right] \in {\mathbb{C}}^{N\!_{\rm A} \times K}$.
As a result of ${{\bf{H}}_{\rm{\!A\!C\!A}}}$ in~\eqref{ChannelACA}, serious DIRS-based ACA interference (a type of IUI) is introduced. To quantify this DIRS-based ACA interference, the SJNR for LU$_k$ denoted by ${\eta _k}$ can be defined with reference to the definition of the signal-to-leakage-plus-noise ratio~\cite{RefSLNRadd}. Specifically, based on~\eqref{RecSig}, ${\eta _k}$ is given by
\begin{equation}
{\eta _k} = \frac{{\mathbb{E}}\!\!\left[{{{\left| {\boldsymbol{h}\!_{D\!T,k}^H{\boldsymbol{w}_{{\rm{\!Z\!F}},k}}} \right|}^2}}\right]}{{\sum\limits_{u \ne k} \!{\mathbb{E}}\!\!\left[{{{\left| {\boldsymbol{h}\!_{D\!T,u}^H{\boldsymbol{w}_{{\rm{\!Z\!F}},k}}} \right|}^2} }\right] }+ {\sigma^2} }.
\label{eqSLNR}
\end{equation}
The ZF precoder in~\eqref{ZF} is calculated based on the CSI of ${\bf{H}}_{\!R\!P\!T}$ and is then fixed during the \emph{DT} phase. Consequently, for a given channel coherence interval, ${\eta _k}$ in~\eqref{eqSLNR} reduces to
\begin{equation}
{\eta _k}  =  \frac{   {{{  {\boldsymbol{w}_{{\rm{ ZF}},k}^H}{\mathbb{E}}\!\left[{\boldsymbol{h}\!_{D\!T,k}}{\boldsymbol{h}\!_{D\!T,k}^H}\right]\!{\boldsymbol{w}_{{\rm{ZF}},k}}   } }}  }{ {\sum\limits_{u \ne k}\! {{{  {\boldsymbol{w}_{{\rm{ ZF}},k
}^H}{\mathbb{E}}\!\left[{\boldsymbol{h}\!_{D\!T,u}}{\boldsymbol{h}\!_{D\!T,u}^H}\right]\!{\boldsymbol{w}_{{\rm{ZF}},k}}   } }}  }\!\!+\! {\sigma^2} }.
\label{eqSLNRred}
\end{equation}

In a traditional MU-MISO system, all channels involved are assumed to be unchanged during the channel coherence time, i.e., ${\bf{H}}_{\!R\!P\!T} = {\bf{H}}_{\!D\!T}$ and ${\bf{H}}_{\rm{\!A\!C\!A}} = {\bf 0}$. Using the ZF precoder, the term ${\mathscr{I}} = {\sum\nolimits_{u \ne k}\! {{{  {\boldsymbol{w}_{{\rm{ ZF}},k
}^H}{\mathbb{E}}\!\left[{\boldsymbol{h}\!_{D\!T,u}}{\boldsymbol{h}\!_{D\!T,u}^H}\right]\!{\boldsymbol{w}_{{\rm{ZF}},k}} } }}  }$ in~\eqref{eqSLNRred} would reduce to ${\mathscr{I}} = {\sum\nolimits_{u \ne k}\! {{{ \left|{\boldsymbol{h}_{\!R\!P\!T,u}^H}  {\boldsymbol{w}_{{\rm{ZF}},k}} \right|^2   } }}  } = 0$. However, due to the DIRS-based ACA interference, the term ${\mathscr{I}}$ is no longer equal to zero in an MU-MISO system jammed by a DIRS-based FPJ. Namely, the LUs are jammed by this DIRS-based ACA interference.
\subsection{Channel Model}\label{ChanMod}
In this section, we present the models of all channels involved, i.e., models for ${\bf G}$, ${{\bf H}}_{\rm I}$, and ${{\bf H}}_{\rm d}$. Specifically, the overall direct channel ${{\bf H}}_{\rm d}$ and the overall DIRS-LU channel ${{\bf H}}_{\rm I}$ are constructed based on the far-field model~\cite{BookFarFeild}. Mathematically, ${{\bf H}}_{\rm I}$ and  ${{\bf H}}_{\rm d}$ are given by
\begin{alignat}{1}
&{{\bf{H}}_{\rm{I}}} = {\widehat {\bf{H}}_{\rm{I}}}{{\bf{D}}_{\rm I}^{{1 \mathord{\left/
 {\vphantom {1 2}} \right.
 \kern-\nulldelimiterspace} 2}}} = \left[ {{\sqrt{{{\mathscr{L}}_{{\rm I},1}}}}{{\widehat {\boldsymbol{h}}}_{{\rm I},1}},  \cdots ,{\sqrt{{{\mathscr{L}}_{{\rm I},K}}}}{{\widehat {\boldsymbol{h}}}_{{\rm I},K}}} \right], \label{HIkeq}\\
&{{\bf{H}}_{\rm{d}}} = {\widehat {\bf{H}}_{\rm{d}}}{{\bf{D}}_{\rm d}^{{1 \mathord{\left/
 {\vphantom {1 2}} \right.
 \kern-\nulldelimiterspace} 2}}} = \left[ {{\sqrt{{{\mathscr{L}}_{{\rm d},1}}}}{{\widehat {\boldsymbol{h}}}_{{\rm d},1}}, \cdots ,{\sqrt{{{\mathscr{L}}_{{\rm d},K}}}}{{\widehat {\boldsymbol{h}}}_{{\rm d},K}}} \right],
\label{Hdkeq}
\end{alignat}
where the elements of the $K\times K$ diagonal matrices ${\bf{D}}_{\rm I} = {\rm{diag}}\left({{\mathscr{L}}_{{\rm I},1}},{{\mathscr{L}}_{{\rm I},2}},\cdots,{{\mathscr{L}}_{{\rm I},K}}\right)$ and ${\bf{D}}_{\rm d} = {\rm{diag}}\left({{\mathscr{L}}_{{\rm d},1}},{{\mathscr{L}}_{{\rm d},2}},\cdots,{{\mathscr{L}}_{{\rm d},K}}\right)$ denote the large-scale channel fading coefficients, which are assumed to be independent~\cite{LDetector1}. The elements of ${\widehat {\bf{H}}_{\rm{I}}}$ and ${\widehat {\bf{H}}_{\rm{d}}}$ are assumed to be i.i.d. Gaussian random variables~\cite{BookFarFeild} defined as $\left[{\widehat {\bf{H}}_{\rm{I}}}\right]_{r,k},\left[{\widehat {\bf{H}}_{\rm{d}}}\right]_{n,k} \sim \mathcal{CN}\left(0,1\right), r=1,2,\cdots, N_{\rm D}$, $n=1,2,\cdots, N_{\rm A}$, and $k = 1,2,\cdots,K$.

The DIRS is assumed to be deployed near the legitimate AP to maximize the jamming impact, and it needs to be equipped with a large number of reflecting elements to launch a significant fully-passive jamming attack since the cascaded large-scale channel fading in the DIRS-jammed channel is much more severe than the fading in the overall LU direct channel~\cite{RIS256ele}. 
Therefore, the AP-DIRS channel $\bf G$ 
is constructed based on the near-field model~\cite{NearfieldMo,NearfieldMo1}:
\begin{equation}
    \begin{split}
    {{\bf{G}}} = {\sqrt{{\mathscr{L}}_{\rm G}}}  \left(  {{\widehat{\bf{G}}}^{{\rm{LOS}}}}{\sqrt {{{{\boldsymbol{\mathscr{Z}}}}}{\left({{{\boldsymbol{\mathscr{Z}}}}} + {\bf{I}}_{N\!_{\rm A}}\right)^{-1}} }} \right.\\
    \left. +~ {{\widehat {\bf{G}}}^{{\rm{NLOS}}}}{\sqrt {{\left({{{\boldsymbol{\mathscr{Z}}}}}+{\bf{I}}_{N\!_{\rm A}}\right)^{-1}}}} \right),
    \end{split}
\label{Ricianchan}
\end{equation}
where ${\mathscr{L}}_{\rm G}$ denotes the large-scale channel fading between 
the AP and the DIRS, the diagonal matrix
 ${\boldsymbol{\mathscr{Z}}}={\rm{diag}}\left({\varepsilon _1},{\varepsilon _2},\cdots,{\varepsilon _{N_{\rm A}}}\right) \in \mathbb{C}^{N\!_{\rm{A}} \times  N\!_{\rm{A}}}$ consists of the Rician factors, 
 and each Rician factor is the ratio of signal power 
 in the line-of-sight (LOS) component to the scattered power 
 in the non-line-of-sight (NLOS) component. 
 The NLOS component ${{\widehat {\bf{G}}}^{{\rm{NLOS}}}}$ is also assumed to follow Rayleigh fading, with elements that satisfy $\left[{{\widehat {\bf{G}}}^{{\rm{NLOS}}}}\right]_{r,n} \sim \mathcal{CN}\left(0,1\right), r=1,2,\cdots, N_{\rm D}$ and $n=1,2,\cdots, N_{\rm A}$.
The elements $\left[{{\widehat {\bf{G}}}^{{\rm{LOS}}}}\right]_{r,n}$ of 
the LOS component ${{\widehat {\bf{G}}}^{{\rm{LOS}}}}$ are given by~\cite{MyGC23,NearfieldMo1}
\begin{equation}
\left[{\widehat {\bf{G}}}^{{\rm{LOS}}}\right]_{r,n} = {e^{ - j\frac{{2\pi }}{\lambda }\left( {{D_n^r} - {D_n}} \right)}},
\label{GLOS}
\end{equation}
where $\lambda$ denotes the wavelength of the transmit signals, and ${D_n^r}$ and ${D_n}$ represent the distance between the $n$-th antenna and the $r$-th DIRS reflecting element, and the distance between the $n$-th antenna and the centre (origin) of the DIRS, respectively. Moreover, the distances between two adjacent DIRS reflecting elements and two adjacent transmit antennas are assumed to be $d ={\lambda}/2$. We identify the locations of the 1-st antenna and DIRS reflecting element as the deployment locations of the legitimate AP and the DIRS.
\subsection{Preliminary: Review of Some Related Results}\label{RandomVar}
\subsubsection{Lindeberg-L$\acute{e}$vy Central Limit Theorem}
Suppose ${\boldsymbol{x}} \buildrel \Delta \over = {\left[ {{x_1},{x_2}, \cdots ,{x_n}} \right]}$ is a vector of i.i.d. random variables with mean ${\mathbb E}\left[x_1\right]= {\mathbb E}\left[x_2\right] = \cdots = {\mathbb E}\left[x_n\right] = \mu < \infty$ and variance ${\rm {Var}}\left[x_1\right]= {\rm {Var}}\left[x_2\right] = \cdots = {\rm {Var}}\left[x_n\right] = {\nu}^2 < \infty$. According to the Lindeberg-L$\acute{e}$vy central limit theorem, the random variable $\sqrt n \left( {\overline X  - \mu } \right)$ converges in distribution to $\mathcal{CN}\left( {0,{\nu^2}} \right)$ as $n \to \infty$, i.e.,
\begin{equation}
\sqrt n \left( {\overline X  - \mu } \right) = \frac{{\sum\limits_{i = 1}^n {{x_i}} }}{{\sqrt n }} - \sqrt n \mu \mathop  \to \limits^{\rm{d}} \mathcal{CN}\left( {0,{\nu ^2}} \right),\;{\rm{as}}\;n \to \infty.
\label{CLTeq}
\end{equation}
\subsubsection{Generalized Rayleigh Quotient Result}\label{Rayleigh}
For a fixed symmetric matrix ${\bf A}\in {\mathbb{C}}^{n\times n}$, the normalized quadratic form $\frac{{{\boldsymbol{x}}^H}\!{\bf A}{\boldsymbol{x}}}{{{\boldsymbol{x}}^H}{\boldsymbol{x}}}$ is referred to as Rayleigh quotient. Furthermore, given a positive definite matrix ${\bf B}\in {\mathbb{C}}^{n\times n}$, the quantity $\frac{{{\boldsymbol{x}}^H}\!{\bf A}{\boldsymbol{x}}}{{{\boldsymbol{x}}^H}\!{\bf B} {\boldsymbol{x}}}$ is called a generalized Rayleigh quotient. The generalized Rayleigh quotient satisfies the following property~\cite{RayleighRes}:
\begin{equation}
\frac{{{\boldsymbol{x}}^H}\!{\bf A}{\boldsymbol{x}}}{{{\boldsymbol{x}}^H}\!{\bf B} {\boldsymbol{x}}} \le {{\lambda _{\max }}\!\left( {{\bf{A}},{\bf{B}}} \right)},
\label{MaxGeneig}
\end{equation}
where ${{\lambda _{\max }}\!\left( {{\bf{A}},{\bf{B}}} \right)}$ is the maximum generalized eigenvalue of $\bf A$ and $\bf B$. The equality in~\eqref{MaxGeneig} holds if and only if ${\boldsymbol{x}} = {\rm{max}}. {\rm{gen.}} {\rm{eigenvector}} \left({\bf A}, {\bf B}\right)$ and ${\rm{max}}. {\rm{gen.}} {\rm{eigenvector}} \left({\bf A}, {\bf B}\right)$ denotes the generalized eigenvector of $\bf A$ and $\bf B$ associated with ${{\lambda _{\max }}\!\left( {{\bf{A}},{\bf{B}}} \right)}$. More specifically, ${{\lambda _{\max }}\left( {{\bf{A}},{\bf{B}}} \right)}$ is given by ${\lambda _{\max }}\!\left( {{\bf{A}},{\bf{B}}} \right) = \max \lambda \!\left( {{\bf{A}},{\bf{B}}} \right)$, where
\begin{equation}
\lambda \!\left( {{\bf{A}},{\bf{B}}} \right) = \left\{ {\left. \lambda  \right|\det \left( {{\bf{A}} - \lambda {\bf{B}}} \right) = 0 } \right\}.
\label{MaxGeneiger}
\end{equation}
If ${\bf{B}}$ is an invertible matrix, the following equation can be further obtained
\begin{equation}
{\rm{max}}. {\rm{gen.}} {\rm{eigenvector}} \left({\bf A}, {\bf B}\right) = {\rm{max}}.{\rm{eigenvector}} \left({\bf B}^{-1}{\bf A}\right),
\label{MaxGeneigvec}
\end{equation}
where ${\rm{max}}.{\rm{eigenvector}} \left({\bf B}^{-1}{\bf A}\right)$ represents the eigenvector of the matrix $\left({\bf B}^{-1}{\bf A}\right)$ associated with the largest eigenvalue.

\section{Anti-Jamming Precoding Against Disco-IRS-Based Fully-Passive Jammers}\label{AntiJamming}
In this section, we first derive the statistical characteristics of the DIRS-based ACA channels for both the persistent DIRS-based FPJ in Section~\ref{AntiPrecoder} and the temporal DIRS-based FPJ. Based on the derived statistical characteristics, we further develop an anti-jamming precoder and prove that it can achieve the maximum SJNR. In Section~\ref{AntiFrame}, we develop a data frame structure that the legitimate AP can use to estimate the statistical characteristics without changing either the legitimate AP architecture (e.g., no additional hardware) or cooperating with the illegitimate DIRS. Furthermore, we explain the mechanism by which it works.

\subsection{Anti-Jamming Precoding Based on Statistical Characteristics of DIRS-Based ACA Channels}\label{AntiPrecoder}
According to the SJNR optimization metric in~\eqref{eqSLNRred}, the fully-passive jamming attacks are caused by the DIRS-based ACA channel ${\bf{H}}_{\rm {\!A\!C\!A}}$ denoted by~\eqref{ChannelACA}.
However, it is unrealistic to acquire the CSI of ${\bf{H}}_{\rm {\!A\!C\!A}}$ unless the ACA is introduced based on the scheme in~\cite{DIRSVT}, i.e., the DIRS phase shifts change only once during the \emph{DT} phase.
In the persistent DIRS-based FPJ here and the temporal DIRS-based FPJ in~\cite{MyGC23,DIRSTWC}, this solution does not work since the DIRS phase shifts change multiple times during the \emph{DT} phase.
Although the legitimate AP can jointly retrain the overall channel ${\bf H}_{\!D\!T}$ with the LUs in the \emph{DT} phase, it can not acquire the useful ${\bf{H}}_{\rm {\!A\!C\!A}}$ by computing $\left( {\bf H}_{\!D\!T} - {\bf{H}}_{ {\!R\!P\!T}} \right)$.

As described in Section~\ref{DIRSFPJ}, the period of the time-varying DIRS reflecting vector $\boldsymbol{\varphi}\!\left( t \right)$ is about the length of the \emph{RPT} phase $T_R$. In other words, the DIRS rapidly ages the wireless channels, and the channel coherence time $T_C$ is shortened to approximately $T_R$. To obtain the useful ${\bf{H}}_{\rm {\!A\!C\!A}}$, the legitimate AP would need to train for the overall channel ${\bf H}_{\!D\!T}$ with a period of $T_R$, and there would be no time available for data transmission.

In summary, the legitimate AP is only able to use the statistical characteristics of ${\bf{H}}_{\rm {\!A\!C\!A}}$ to design an anti-jamming precoder against the persistent DIRS-based FPJ presented in Section~\ref{DIRSFPJ}.
In order to develop a practical anti-jamming precoder, therefore, we derive the following statistical characteristics of ${\bf{H}}_{\rm {\!A\!C\!A}}$.

\newtheorem{proposition}{Proposition}
\begin{proposition}
    \label{Proposition1}
    The i.i.d. elements of ${\bf{H}}_{\rm {\!A\!C\!A}}$ converge in distribution to $\mathcal{CN}\!\left( {0,  {{{\mathscr{L}}\!_{{\rm G}}}{{\mathscr{L}}\!_{{\rm I},k}}{N\!_{\rm D}} {\overline \alpha} } } \right)$ as $N_{\rm D} \to \infty$, i.e.,
    \begin{equation}
        {\left[ {{\bf H}_{\rm {\!A\!C\!A}}} \right]_{k,n}} \mathop  \to \limits^{\rm{d}} \mathcal{CN}\!\left( {0,  {{{\mathscr{L}}\!_{{\rm G}}}{{\mathscr{L}}\!_{{\rm I},k}}{N\!_{\rm D}}{\overline \alpha} } } \right), \forall k,n,
        \label{HDSta}
    \end{equation}
    where 
    $\overline \alpha = \sum\nolimits_{i1 = 1}^{{2^b}} \sum\nolimits_{i2 = 1}^{{2^b}}  {P_{i1}}{P_{i2}} \big( {\mu _{i1}^2} + {\mu _{i2}^2} - 2{\mu _{i1}}{\mu _{i2}}$ $\cos ( {{\theta _{i1}} - {\theta _{i2}}} ) \big)$,
    ${\mu _{i1}}$, ${\mu _{i2}} \in \Omega$, and ${\theta _{i1}}$, ${\theta _{i2}} \in \Theta$.
\end{proposition}

\begin{IEEEproof}
    See Appendix~\ref{AppendixA}.
\end{IEEEproof}

On the other hand, if the DIRS only changes its reflection coefficients during the \emph{DT} phase and remains silent$\footnote{The term ``silent" means that the wireless signals are perfectly absorbed by the DIRS, which can be achieved by setting the illegitimate IRS in a special  mode~\cite{RISSilent}.}$ during the \emph{RPT} phase as in~\cite{MyGC23,DIRSTWC}, the statistical characteristics of ${\bf{H}}_{\rm {\!A\!C\!A}}$ change, as shown in Proposition~\ref{Proposition2}.

\begin{proposition}
    \label{Proposition2}
    The i.i.d. elements of ${\bf{H}}_{\rm {\!A\!C\!A}}$ converge in distribution to $\mathcal{CN}\!\left( {0,  {{{\mathscr{L}}\!_{{\rm G}}}{{\mathscr{L}}\!_{{\rm I},k}}{N\!_{\rm D}} {\overline \alpha} } } \right)$ as $N_{\rm D} \to \infty$, i.e.,
    \begin{equation}
        {\left[ {{\bf H}_{\rm {\!A\!C\!A}}} \right]_{k,n}} \mathop  \to \limits^{\rm{d}} \mathcal{CN}\!\left( {0,  {{{\mathscr{L}}\!_{{\rm G}}}{{\mathscr{L}}\!_{{\rm I},k}}{N\!_{\rm D}}{\overline \alpha} } } \right), \forall k,n,
        \label{HDStaxx}
    \end{equation}
    where $\overline \alpha = {{\sum\nolimits_{i = 1}^{{2^b}} {P_i}{\mu_i^2} }}$.
\end{proposition}

\begin{IEEEproof}
    See Appendix~\ref{AppendixB}.
\end{IEEEproof}

Based on Propositions~\ref{Proposition1} and~\ref{Proposition2}, the statistical characteristics of the DIRS-based ACA channel depend on the distribution of DIRS phase shifts when the gains of the DIRS reflecting elements are a function of the corresponding phase shifts. Furthermore, the jamming impact of the persistent DIRS-based FPJ is different from that of the temporal DIRS-based FPJ in~\cite{MyGC23,DIRSTWC}.
According to Propositions~\ref{Proposition1} and~\ref{Proposition2}, the SJNR for LU$_k$ in~\eqref{eqSLNRred} reduces to
\begin{equation}
    {\eta _k} \!\!=\!\! \frac{ {{{\left| {\boldsymbol{h}_{{\!R\!P\!T},k}^H{\boldsymbol{w}\!_{{\rm{ZF}},k}}} \right|}^2}} \!\! + \! {{{  {\boldsymbol{w}_{{\rm{ZF}},k}^H}{\mathbb{E}}\!\left[\boldsymbol{h}_{{\rm{\!A\!C\!A}},k}{\boldsymbol{h}_{{\rm{\!A\!C\!A}},k}^{H}}\right]\!{\boldsymbol{w}\!_{{\rm{ZF}},k}}   } }}  }{\!{\sum\limits_{u \ne k} \!\!\left(\! {{{\left| {\boldsymbol{h}\!_{{\!R\!P\!T},u}^H{\!\boldsymbol{w}\!_{{\rm{ZF}},k}}} \right|}^2}}\! \!+ \!  {{{  {\boldsymbol{w}_{{\rm{ZF}},k}^H}{\mathbb{E}}\!\left[\boldsymbol{h}_{{\rm{\!A\!C\!A}},u}{\boldsymbol{h}_{{\rm{\!A\!C\!A}},u}^{H}}\right]\!{\boldsymbol{w}_{{\rm{ZF}},k}}   } }} \!\!\right) } \!\!+\! {\sigma^2} }.
    \label{eqSLNRredadd}
\end{equation}


In order to suppress the DIRS-based ACA interference, the legitimate AP should employ an anti-jamming precoder to maximize the SJNRs. However, as mentioned in Section~\ref{Intro}, it is unrealistic for the legitimate AP to have CSI for the DIRS-based ACA channel ${\bf{H}}_{\rm {\!A\!C\!A}}$. In other words, the AP can not exploit the CSI of ${\bf{H}}_{\rm {\!A\!C\!A}}$ to design an anti-jamming precoder. To this end, we derive an anti-jamming precoder in Theorem~\ref{Theorem1} that can maximize the SJNR expressed by~\eqref{eqSLNRredExp}.
\newtheorem{theorem}{Theorem}
\begin{theorem}
\label{Theorem1}
The optimal anti-jamming precoder for LU$_k$ to mitigate the DIRS-based fully-passive jamming attacks, i.e., to maximize the SJNR $\eta_k$, is given by
\begin{equation}
{\boldsymbol{w}\!_{{\rm{Anti}},k}} \propto \max.{\rm{eigenvector}}\left({\bf A}_k\right),
\label{AntiJamm}
\end{equation}
where
\begin{alignat}{1}
\nonumber
{\bf A}_k = &  \left({{\boldsymbol{h}_{{\!R\!P\!T},k}}\boldsymbol{h}_{{{\!R\!P\!T}},k}^H + {{{\mathscr{L}}\!_{{\rm G}}}{{\mathscr{L}}\!_{{\rm I},k}}{N\!_{\rm D}} {\overline \alpha} }{{\bf{I}}\!_{N\!_{\rm A}}}}\right) \times\\
&\;\;\left(\!\!{{ {{{\widetilde {\bf{H}}}_{{\!R\!P\!T},k}}\widetilde {\bf{H}}_{{\!R\!P\!T},k}^H + \!\!\left( \!\!{\frac{{{\sigma ^2}}K}{{{P_0}}} \!+ \!\! {\sum\limits_{u \ne k}\!{{\mathscr{L}}\!_{{\rm G}}}{{\mathscr{L}}\!_{{\rm I},u}}{N\!_{\rm D}} {\overline \alpha} }}\! \right)\!{{\bf{I}}\!_{N\!_{\rm A}}}}  }}\!\!\right)^{\!\!{-1}},
\label{AntiJammMat}
\end{alignat}
and ${{\widetilde {\bf{H}}}_{{\!R\!P\!T},k}} \!=\! \left[{\boldsymbol{h}_{{\!R\!P\!T},1}},\cdots,{\boldsymbol{h}_{{\!R\!P\!T},k-1}},{\boldsymbol{h}_{{\!R\!P\!T},k+1}},\cdots,{\boldsymbol{h}_{{\!R\!P\!T},K}} \right]$.
\end{theorem}

\begin{IEEEproof}
According to Propositions~\ref{Proposition1} and~\ref{Proposition2}, we can rewrite ${\eta _k}$ in~\eqref{eqSLNRredadd} to
\begin{equation}
{\eta _k}  \!\!=\!\!  \frac{   {{{  {\boldsymbol{w}_{{\rm{ZF}},k}^H} \!\left({\boldsymbol{h}\!_{R\!P\!T,k}}{\boldsymbol{h}\!_{R\!P\!T,k}^H} {+} {{{\mathscr{L}}\!_{{\rm G}}}{{\mathscr{L}}\!_{{\rm I},k}}{N\!_{\rm D}} {\overline \alpha}} {\bf I}_{N\!_{\rm A}}  \right)\!{\boldsymbol{w}_{{\rm{ZF}},k}}   } }}  }{ {\sum\limits_{u \ne k}\! {{{  {\boldsymbol{w}\!_{{\rm{ ZF}},k
}^H}\!\left({\boldsymbol{h}\!_{R\!P\!T,u}}{\boldsymbol{h}\!_{R\!P\!T,u}^H} {+} {{{\mathscr{L}}\!_{{\rm G}}}{{\mathscr{L}}\!_{{\rm I},u}}{N\!_{\rm D}} {\overline \alpha}} {\bf I}_{N\!_{\rm A}} \right)\!{\boldsymbol{w}_{{\rm{ZF}},k}}   } }}  }\!\!+\! {\sigma^2} }.
\label{eqSLNRredExp}
\end{equation}
Furthermore, we rewrite~\eqref{eqSLNRredExp} as:
\begin{equation}
{\eta _k} = \frac{ {{{  {\boldsymbol{w}_{{\rm{ZF}},k}^H} { \widehat {\bf{H}}}\!_{D\!T,k} {{\boldsymbol{w}_{{\rm{ZF}},k}}}  } }} }{ {{{  {\boldsymbol{w}_{{\rm{ZF}},k}^H} { \widehat {\widetilde {\bf H}} }\!_{D\!T,k} {{\boldsymbol{w}_{{\rm{ZF}},k}}}  } }}  },
\label{eqSLNRequMore}
\end{equation}
where ${ \widehat {\bf{H}}} _{\!D\!T,k} \!\!=\!\! {{ { {\boldsymbol{h}_{{\!R\!P\!T},k}} \boldsymbol{h}_{{\!R\!P\!T},k}^H  ~+~  \left( {  { {{\mathscr{L}} _{{\rm G}}}{{\mathscr{L}}_{{\rm I},k}}{N\!_{\rm D}}{\overline \alpha} }} \right){{\bf{I}}\!_{N\!_{\rm A}}}}  }}$, ${ \widehat {\widetilde {\bf H}} }\!_{D\!T,k} \!=\!  {{ {{{\widetilde {\bf{H}}}_{{\!R\!P\!T},k}} \widetilde {\bf{H}}_{{\!R\!P\!T},k}^H \!+ \!\left(\! \!{\frac{{{\sigma ^2}}K}{{{P_0}}} \!+\! {\!\!\sum\limits_{u \ne k}\!{{\mathscr{L}}\!_{{\rm G}}}{{\mathscr{L}}\!_{{\rm I},u}}{N\!_{\rm D}}{\overline \alpha}}} \!\right)\!\!{{\bf{I}}\!_{N\!_{\rm A}}}}  }}$, and $k=1,2,\cdots,K$.

Using the generalized Rayleigh quotient result, we have
\begin{equation}
{\eta _k} \le {\lambda _{\max }}\!\left(\! { { \widehat {\bf{H}}}\!_{D\!T,k},{ \widehat {\widetilde {\bf H}} }\!_{D\!T,k} } \right).
\label{eqRayRitz}
\end{equation}
When \eqref{eqRayRitz} holds with equality, the maximum SJNR $\eta_k$ is obtained. More specifically, the optimal anti-jamming precoder for LU$_k$ that maximizes $\eta_k$ is given by
\begin{equation}
{\boldsymbol{w}\!_{{\rm{Anti}},k}} = \sqrt{p_k}\frac{\max.{\rm{eigenvector}}\left({\bf A}_k\right)}{\|\max.{\rm{eigenvector}}\left({\bf A}_k\right)\|},
\label{AntiJammPrecoding}
\end{equation}
where $\eta_k = {\lambda _{\max }}\!\left(\! { { \widehat {\bf{H}}}\!_{D\!T,k},{ \widehat {\widetilde {\bf H}} }\!_{D\!T,k} } \right)$.
\end{IEEEproof}
\subsection{Frame Design for Obtaining Statistical Characteristics of DIRS-Based ACA Channels}\label{AntiFrame}
Theorem~\ref{Theorem1} presented an anti-jamming precoder and proved that it can maximize the SJNRs for the LUs. The designed anti-jamming precoder requires only the statistical characteristics of the DIRS-based ACA channel ${\bf H}_{\rm{\!A\!C\!A}}$ which were derived in Propositions~\ref{Proposition1} and~\ref{Proposition2}. In this section, we explain how in practice the legitimate AP can acquire the statistical characteristics without changing its architecture (e.g., additional hardware and operating procedure) or cooperating with the illegitimate DIRS.
\begin{figure}[!t]
    \centering
    \includegraphics[scale=0.85]{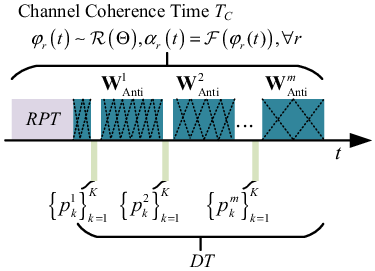}
    \caption{A data frame structure for the legitimate AP to estimate the statistical characteristics of DIRS-based ACA channels.}
    \label{fig3}
\end{figure}

Fig.~\ref{fig3} illustrates the data frame structure that can be used for the legitimate AP to estimate the statistical characteristics of ${\bf H}_{\rm{\!A\!C\!A}}$. 
Within a channel coherence time, when the LUs perceive that they are being jammed (e.g., due to a significant performance degradation), they feed back their received power to the legitimate AP (e.g.,~\cite{JammResilientCommun}). 
Note that only a few bits are required to feed back the received power values to the AP. 
We denote the $s$-th set of feedback for the LU received power as $\left\{ {p_k^s} \right\}_{k = 1}^K$ $(1 \le s \le m \le C)$, as depicted in Fig.~\ref{fig3}, where $m$ is the total number of feedback transmissions and $C$ is the ratio of the length of the \emph{DT} phase to the \emph{RPT} phase.
Based on all $s$ sets of feedback, we compute the $s$-th estimate of the statistical characteristic of ${{{\boldsymbol{h}}_{{\rm{\!A\!C\!A}},k}}}$ by
\begin{equation}
    {\left. {\overline {{{{\mathscr{L}}\!_{{\rm G}}}{{\mathscr{L}}\!_{{\rm I},k}}{N\!_{\rm D}}{\overline \alpha} }}} \right|_s}   \!=\!
    \frac{\left|{K\sum\limits_{{i_s} = 1}^s {p_k^{i_s}}  - sK\left\| {{{\boldsymbol{h}}_{RPT,k}}} \right\|_{\rm{F}}^2} \right|}{{{P_0}{N_{\rm{A}}}s}}, 1 \!\le\! s \!\le \!m,
    \label{Estimatedstacha}
\end{equation}
where the absolute value ensures that the estimate of ${\left. {\overline {{{{\mathscr{L}}\!_{{\rm G}}}{{\mathscr{L}}\!_{{\rm I},k}}{N\!_{\rm D}}{\overline \alpha} }}} \right|_s} $ is positive.
In~\eqref{Estimatedstacha}, the CSI of ${{{\boldsymbol{h}}_{RPT,k}}}$ was obtained in the \emph{RPT} phase, and thus $\left\| {{{\boldsymbol{h}}_{RPT,k}}} \right\|_{\rm{F}}^2$ is known by the legitimate AP.

The derivation of the estimate in~\eqref{Estimatedstacha} can be found by noting that
\begin{alignat}{1}
    \nonumber
    {\mathbb{E}}\!\left[\left\| {\bf H}_{\!D\!T} \right\|^2\right] &= {\mathbb{E}}\!\left[\left\|  {\bf H}_{\!R\!P\!T}  + {\bf H}_{\rm{\!A\!C\!A}} \right\|^2\right] \\ \nonumber
    &={\mathbb{E}} \Big[ {\rm{tr}} \big(  {\bf H}_{\!R\!P\!T}{\bf H}_{\!R\!P\!T}^H  + {\bf H}_{\rm{\!A\!C\!A}}{\bf H}_{\rm{\!A\!C\!A}}^H \\
    &~~~+{\bf H}_{\rm{\!A\!C\!A}}{\bf H}_{\!R\!P\!T}^H+ {\bf H}_{\!R\!P\!T}{\bf H}_{\rm{\!A\!C\!A}}^H \big) \Big].
    \label{Estimatedstachaexp}
\end{alignat}
${\bf H}_{\!R\!P\!T}$ is constant during the RPT phase but is random due to the randomly chosen DIRS phase shift. Consequently, \eqref{Estimatedstachaexp} reduces to
\begin{alignat}{1}
    \nonumber
    {\mathbb{E}}\!\left[\left\| {\bf H}_{\!D\!T} \right\|^2\right]  = &\left\| {\bf H}_{\!R\!P\!T} \right\|^2+ {\rm{tr}} \big( {\mathbb{E}}\!\left[ {\bf H}_{\rm{\!A\!C\!A}}{\bf H}_{\rm{\!A\!C\!A}}^H \right] \\
    &+ {\mathbb{E}}\!\left[ {\bf H}_{\rm{\!A\!C\!A}} \right]{\bf H}_{\!R\!P\!T}^H  + {\bf H}_{\!R\!P\!T}{\mathbb{E}}\!\left[ {\bf H}_{\rm{\!A\!C\!A}}^H \right] \big).
\label{EstimatedstachaexpRe}
\end{alignat}
Based on Propositions~\ref{Proposition1} and~\ref{Proposition2}, we have that
\begin{equation}
    {\mathbb{E}}\!\left[\left\| {\bf H}_{\!D\!T} \right\|^2\right]  \mathop  \to \limits^{\rm{d}} \left\| {\bf H}_{\!R\!P\!T} \right\|^2 + {N\!_{\rm A}} \! \sum_k^K{\!{{\mathscr{L}}\!_{{\rm G}}}{{\mathscr{L}}\!_{{\rm I},k}}{N\!_{\rm D}}{\overline \alpha} }.
\label{HACAestSta}
\end{equation}
Furthermore,
\begin{equation}
{{\mathscr{L}}\!_{{\rm G}}}{{\mathscr{L}}\!_{{\rm I},k}}{N\!_{\rm D}}{\overline \alpha} \mathop  \to \limits^{\rm{d}} \frac{{\mathbb{E}}\!\left[\left\| {\boldsymbol{h}}_{\!D\!T,k} \right\|^2\right] - \left\| {\boldsymbol{h}}_{\!R\!P\!T,k} \right\|^2}{N\!_{\rm A}} .
\label{HACAestStak}
\end{equation}

According to~\eqref{HACAestStak}, we can compute the $s$-th estimate of 
${\left. {\overline {{{{\mathscr{L}}\!_{{\rm G}}}{{\mathscr{L}}\!_{{\rm I},k}}{N\!_{\rm D}}{\overline \alpha} }}} \right|_s}$ 
using~\eqref{Estimatedstacha}, where the expectation ${\mathbb{E}}\!\left[\left\| {\boldsymbol{h}}_{\!D\!T,k} \right\|^2\right]$ is approximated by ${\sum\nolimits_{{i_s} = 1}^s {p_k^{i_s}}}/{s}$.
Substituting~\eqref{Estimatedstacha} into~\eqref{AntiJamm}, the anti-jamming precoder 
${\bf W}_{\rm{\!Anti}}^{s}$ can be calculated. In the following section, 
we illustrate the difference between the theoretical SJNR 
${  { {{\mathscr{L}}\!_{{\rm G}}}{{\mathscr{L}}\!_{{\rm I},k}}{ N\!_{\rm D}}{\overline \alpha} }}$ 
and the estimated value ${\left. {\overline {{{{\mathscr{L}}\!_{{\rm G}}}{{\mathscr{L}}\!_{{\rm I},k}}{N\!_{\rm D}}{\overline \alpha} }}} \right|_s}$ 
in~\eqref{Estimatedstacha}.

\begin{table*}
    \footnotesize
    \centering
    \caption{Wireless Channel Simulation Parameters}
    \label{tab1}
    \begin{threeparttable}
    \begin{tabular}{ c|c|c }
    \hline
    Parameter        &Notation     &Value\\
    \hline
    Large-scale fading of LOS channels     &${ {\mathscr{L}_{\rm{G}}} }$   & $ 35.6 + 22{\log _{10}}({d_{\rm p}}) $ (dB) \\
    \hline
    Large-scale fading of NLOS channels    &${ {\mathscr{L}_{{\rm{d}},k}},{\mathscr{L}_{{\rm{I}},k}}}$    &$32.6+36.7{\log _{10}}({d_{\rm p}})$ \\
    \hline
    Transmission bandwidth                       &$BW$                                                  &180 kHz\\
    \hline
    Rician factors                       &${\boldsymbol{\mathscr{Z}}}$                                 &$10{\bf{I}}_{N\!_{\rm A}}$ \\
    \hline
    Transmission wavelength                       &$\lambda$                                                  &0.05 m\\
    \hline
    Ratio of $T_D$ and $T_R$                      &$C$                                                  &6\\
    \hline
    \end{tabular}
    \end{threeparttable}
\end{table*}

\section{Simulation Results and Discussion}\label{ResDis}
In this section, we present numerical results to determine the feasibility of the anti-jamming precoder given in Section~\ref{AntiJamming} and show the performance of the proposed precoder against both the persistent DIRS-based FPJ in Section~\ref{DIRSFPJ} and the temporal DIRS-based FPJ in~\cite{MyGC23,DIRSTWC}. We assume an MU-MISO system with 12 single-antenna LUs that are jammed by the persistent DIRS-based FPJ and the temporal DIRS-based FPJ, respectively. The legitimate AP has 16 antennas located at (0 m, 0 m, 5 m), and the DIRS with 2048 reflecting elements is deployed at (-$d_{\rm{AD}}$ m, 0 m, 5 m), where the AP-DIRS distance $d_{\rm{AD}}$ is nominally set to 2.
 The LUs are randomly distributed in a circular region $S$ with a radius of 20 m and a centre of (0 m, 180 m, 0 m). If not otherwise specified, the numbers of LUs, AP antennas, DIRS reflecting elements, as well as the AP-DIRS distance in this section default to the values above, i.e., $K=12$, $N_{\rm A}=16$, $N_{\rm D} = 2048$, and $d_{\rm{A\!D}} = 2$.
\begin{figure}[!h]
    \centering
    \includegraphics[scale=0.8]{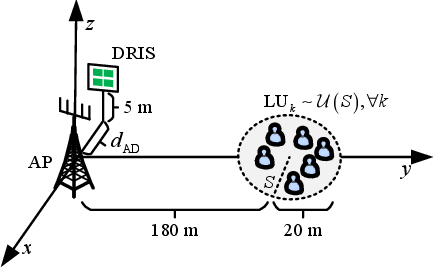}
    \caption{An example of an MU-MISO system jammed by a DIRS-based FPJ, where the LUs are randomly located in the circular region $S$ with a radius of 20 m and a centre of (0 m, 180 m, 0 m) with the same probability, and the legitimate AP and the DIRS are deployed at (0 m, 0 m, 5m) and (-$d_{\rm{AD}}$ m, 0 m, 5m) respectively.}
    \label{fig4}
    \end{figure}
    
The propagation parameters of wireless channels  ${\bf{H}_{\rm D}}$, ${\bf{H}_{\rm I}}$, and ${{\bf{G}}}$  
are given in Table~\ref{tab1}, and are based on standard 3GPP propagation models~\cite{3GPP}. 
The variance of the AWGN noise is $\sigma^2 \!=\!-170\!+\!10\log _{10}\left(BW\right)$ dBm. 
In addition, the wavelength of the transmit signals is assumed to be $\lambda = 0.05$ m. 
The length of the $DT$ phase is 6 times longer than that of the $RPT$ phase, i.e., $T_D = 6T_R$. 
In the following discussion, we will show that the anti-jamming precoder proposed 
in Section~\ref{AntiPrecoder} works for any $m$ in~\eqref{Estimatedstacha}, even for $m=1$.

We assume that the DIRS has one-bit control with phase shift and gain values taken from $\Theta = \{\frac{\pi}{9},\frac{7\pi}{6}\}$ and $\Omega = {\cal F}\!\left({\Theta}\right) = \{0.8,1\}$~\cite{IRSsur1}. Such a design is relatively simple to implement on a massive scale~\cite{IRSsur2}. Based on Propositions~\ref{Proposition1} and~\ref{Proposition2}, the jamming impacts of the persistent DIRS-based FPJ and the temporal DIRS-based FPJ are related to the distribution of the random DIRS phase shifts when the gain value of the $r$-th DIRS reflecting element ${\alpha _r}\!\left( t \right)$ is a function of the corresponding phase shift ${\varphi _r}\!\left( t \right)$. Note that this conclusion is different from the conclusion based on the ideal IRS model in~\cite{DIRSTWC}. 
To show the influence of the DIRS phase shift distributions, we consider the two cases in Table~\ref{tab2}.

\begin{table}
    \footnotesize
    \centering
    \caption{DIRS Phase Shift Distributions}
    \label{tab2}
    \begin{threeparttable}
    \begin{tabular}{ |c|c|c|c|c| }
    \hline
    {}   &${\mathbb{P}}\!\left( {\varphi _r}(t) = {\theta _{1}} \right) $        &${\mathbb{P}}\!\left( {\varphi _r}(t) = {\theta _{2}} \right) $
    &$\overline \alpha$ in~\eqref{HDSta}     &$\overline \alpha$ in~\eqref{HDStaxx}\\
    \hline
    \emph{Case 1}    &$0.25$                                                          &0.75
    &1.2059                                     &0.91\\
    \hline
    \emph{Case 2}    &$0.5$                                                          &0.5
    &1.6078                                     &0.82\\
    \hline
    \end{tabular}
    \end{threeparttable}
\end{table}

\subsubsection{Ergodic LU Rate Versus Transmit Power Based on Derived Statistical Characteristics}
\begin{figure}[!h]
    \centering
    \subfloat{
            \includegraphics[scale=0.4]{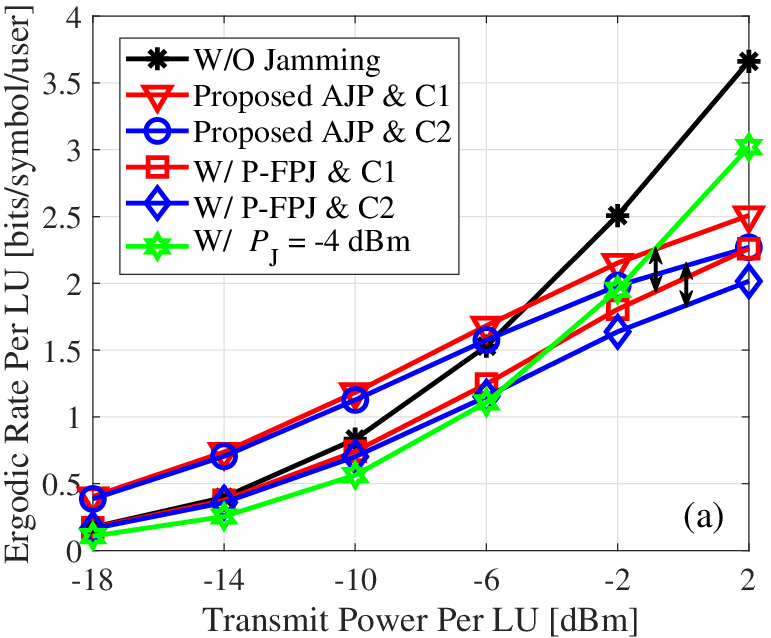}}\\
    \subfloat{
            \includegraphics[scale=0.4]{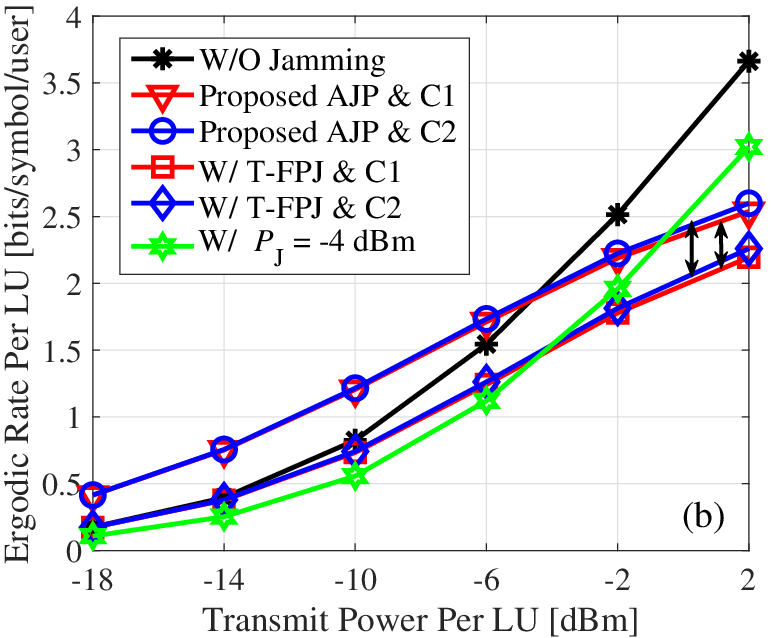}}
   \caption{Ergodic rate of each LU vs average transmit power of each LU for different benchmarks under jamming attacks  launched by (a) the persistent DIRS-based FPJ and (b) the temporal DIRS-based FPJ, respectively.}
    \label{ResFig1}
\end{figure}

To verify the feasibility of the anti-jamming precoder proposed in Section~\ref{AntiPrecoder}, Fig.~\ref{ResFig1} illustrates the relationship between 
the ergodic rate per LU$\footnote{For real-world applications,  we also employ the sum rate $R_{\rm{sum}} = \sum\limits_{k = 1}^K {{{\rm{log}}_2{\!\left(\!\!1+ \!\frac{   {{{\left| {\boldsymbol{h}\!_{D\!T,k}^H{\boldsymbol{w}\!_{k}}} \right|}^2}} }{{\sum\limits_{u \ne k} \!  {{{\left| {\boldsymbol{h}\!_{D\!T,u}^H{\boldsymbol{w}\!_{k}}} \right|}^2} }  }+ {\sigma^2} } \!\right)}}}$ to visualise the performance. Furthermore, the rate per LU is defined as $\frac{R_{\rm{sum}}}{K}$.}$ 
and the transmit power per LU ($\frac{P_0}{K}$) for the persistent DIRS-based FPJ case described in Section~\ref{DIRSFPJ}. The performance of the following  benchmarks is illustrated and compared: the legitimate AP uses the ZF precoder and does not suffer from jamming attacks (W/O Jamming);
the legitimate AP is jammed by the persistent DIRS-based FPJ described in Section~\ref{DIRSFPJ} while the random DIRS phase shifts follow the distribution in \emph{Case 1}  (W/ P-FPJ \& C1) and the second distribution in \emph{Case 2} (W/ P-FPJ \& C2) in Table~\ref{tab2};
the legitimate AP adopts the anti-jamming precoder in Theorem~\ref{Theorem1} for \emph{Case 1} (Proposed AJP \& C1) and \emph{Case 2} (Proposed AJP \& C2);
the legitimate AP suffers from an AJ with -4 dBm jamming power (AJ w/ $P_{\rm J}$ = -4 dBm), where the AJ is deployed at (-2 m, 0 m, 5 m). Fig.~\ref{ResFig1} (b) illustrates the corresponding results for the temporal DIRS-based FPJ case in~\cite{MyGC23,DIRSTWC}.

We see from Fig.~\ref{ResFig1} that the proposed anti-jamming precoder is effective for both the persistent DIRS-based FPJ and the temporal DIRS-based FPJ, and when the transmit power is low it can even achieve a rate higher than the case without any jamming. This is because the proposed anti-jamming precoder can to some extent use the DIRS-based channels to improve its rate per LU. In practice, an MU-MISO system using low-order modulation such as quadrature phase shift keying (QPSK) can operate in the low transmit power domain~\cite{Digitalpbook}.

\begin{figure}[!t]
    \centering
    \subfloat{
            \includegraphics[scale=0.4]{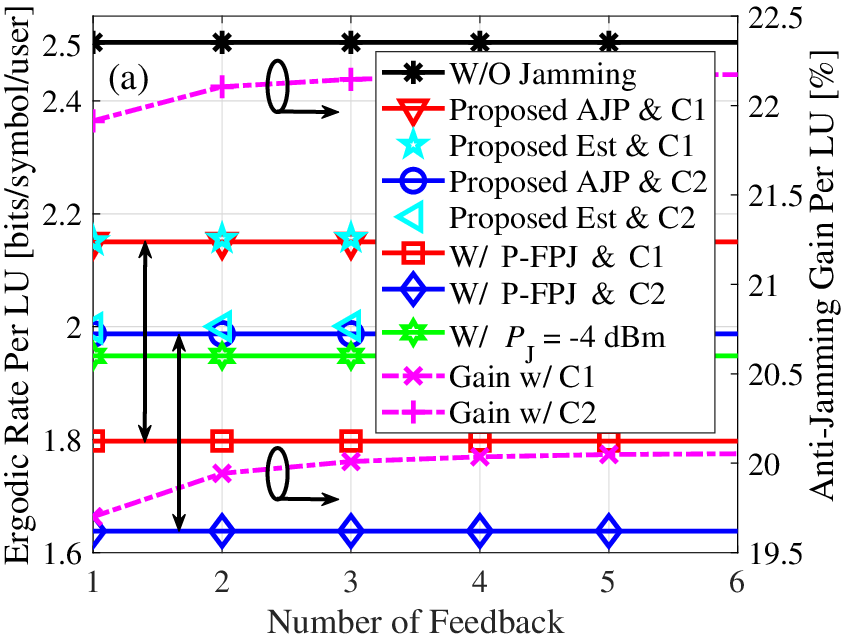}}\\
    \subfloat{
            \includegraphics[scale=0.4]{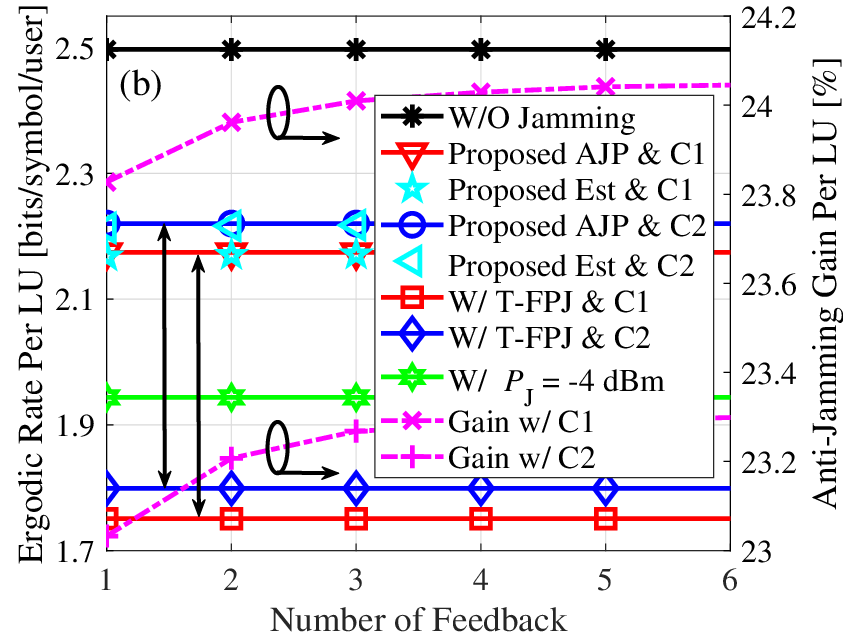}}
   \caption{Relationship between the ergodic rate per LU and the number of feedback transmissions for different benchmarks under jamming attacks launched by (a) the persistent DIRS-based FPJ and (b) the temporal DIRS-based FPJ, where the transmit power per LU is -2 dBm (high transmit power).}
    \label{ResFig2h}
\end{figure}

On the other hand, IUI dominates the noise for high transmit power~\cite{ZFBF}. Although the proposed anti-jamming precoder can to some extent exploit the DIRS-jammed channel to improve the SJNR of each LU, it also amplifies IUI due to the leakage from the DIRS-jammed channel. As a result, the ergodic rate per LU resulting from the anti-jamming precoder is progressively weaker than that without any jamming.
It can be seen that the proposed anti-jamming precoder always mitigates the jamming attacks of both the persistent DIRS-based FPJ and the temporal DIRS-based FPJ. However, these two FPJs exhibit different behaviors for \emph{Case 1} and \emph{Case 2}. This is due to the different $\overline \alpha$ in the two FPJs, as shown in Table~\ref{tab2}. The larger the values of $\overline \alpha$, the more pronounced the jamming effect. Since the two possible values of $\overline \alpha$ in the temporal DIRS-based FPJ are similar to each other, the jamming impacts for \emph{Case 1} and \emph{Case 2} 
are similar, as shown in Fig~\ref{ResFig1}(b). For the persistent DIRS-based FPJ, $\overline \alpha$ in \emph{Case 1} is much smaller than that in \emph{Case 2}. Therefore, the jamming impact of W/ P-FPJ \& C1 is also weaker than that of W/GFPJ \& C2.
Note that the anti-jamming precoder behaves differently in the high and low transmit power domains. Therefore, our following discussions will be focused on the high and lower power cases.
\subsubsection{Ergodic LU Rate Versus Transmit Power Based on Estimated Statistical Characteristics}
Fig.~\ref{ResFig2h} shows the feasibility of the data frame structure presented in Section~\ref{AntiFrame} at -2 dBm transmit power, where the rate per LU resulting from the legitimate AP with the anti-jamming precoder using the estimated statistical characteristics for \emph{Case 1}  and \emph{Case 2} are denoted by Proposed Est \& C1 and Proposed Est \& C2, respectively.
Specifically, Fig.~\ref{ResFig2h}(a) illustrates the results for the persistent DIRS-based FPJ, and Fig.~\ref{ResFig2h}(b) shows the results for the temporal DIRS-based FPJ. It can be seen that the rate per LU based on the 1st estimated statistical characteristic in~\eqref{Estimatedstacha} is good enough. The difference in anti-jamming gain using the 1st estimated statistical characteristic and that using the 6th estimated statistical characteristic is less than 0.5\%. Moreover, there is only a small gap between the ergodic rates calculated with ${{{\mathscr{L}}\!_{{\rm G}}}{{\mathscr{L}}\!_{{\rm I},k}}{N\!_{\rm D}} {\overline \alpha} }$ and ${\left. {\overline {{{{\mathscr{L}}\!_{{\rm G}}}{{\mathscr{L}}\!_{{\rm I},k}}{N\!_{\rm D}}{\overline \alpha} }}} \right|_s}$, which verifies the feasibility of the approach used to estimate the statistical characteristics in Section~\ref{AntiFrame}.

\begin{figure}[!t]
    \centering
    \subfloat{
        \includegraphics[scale=0.4]{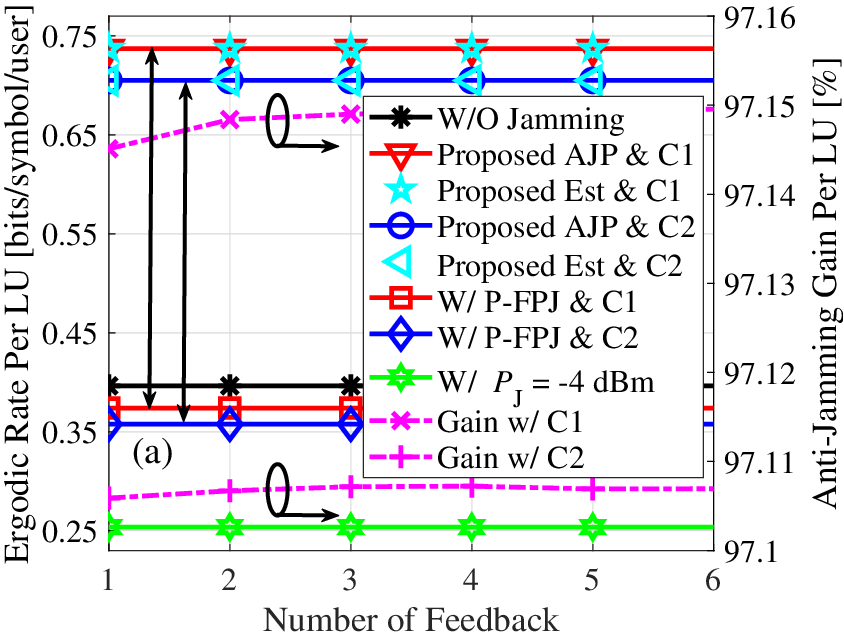}}\\
    \subfloat{
        \includegraphics[scale=0.4]{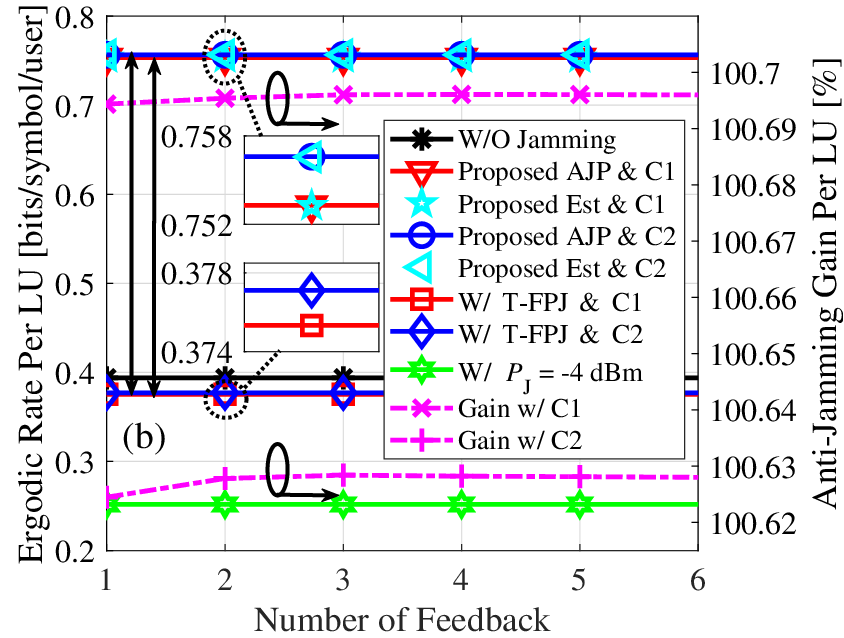}}   
    \caption{Relationship between the ergodic rate per LU and the number of feedback transmissions for different benchmarks under jamming attacks  launched by (a) the persistent DIRS-based FPJ and (b) the temporal DIRS-based FPJ, where the transmit power per LU is -14 dBm (low transmit power).}
	\label{ResFig3l}
\end{figure}

\begin{figure}[!t]
    \centering
    \subfloat{
        \includegraphics[scale=0.4]{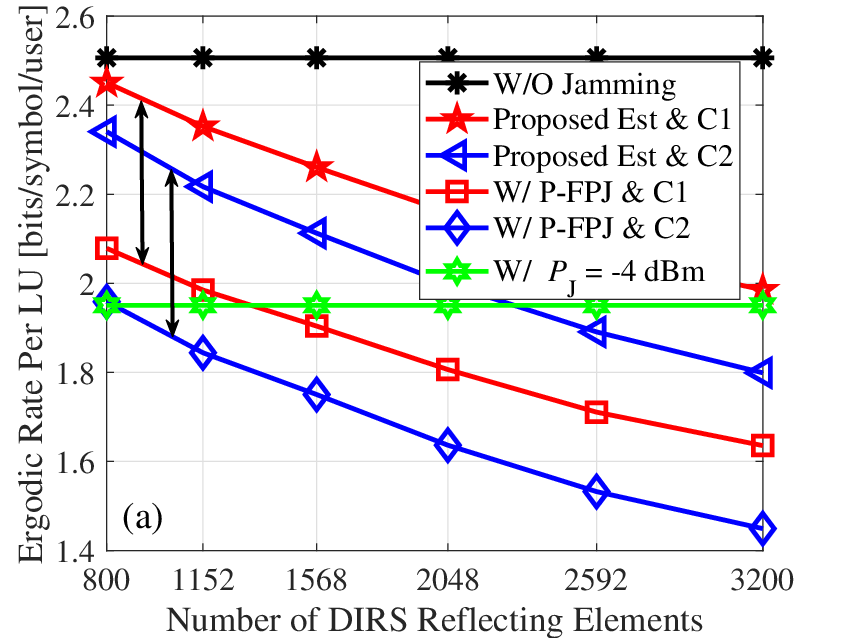}}\\
    \subfloat{
        \includegraphics[scale=0.4]{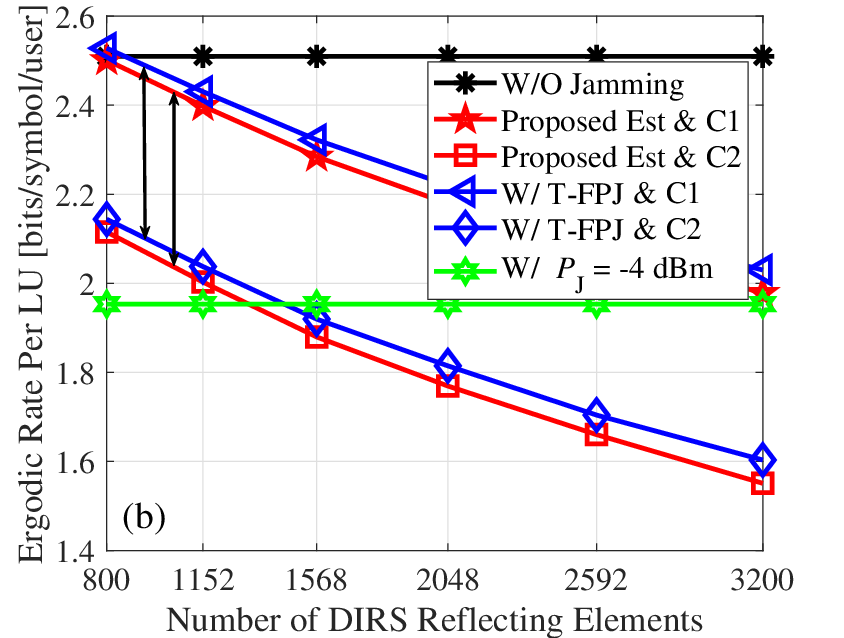}}  
    \caption{Relationship between the ergodic rate per LU and the number of DIRS reflecting elements for different benchmarks under jamming attacks launched by (a) the persistent DIRS-based FPJ and (b) the temporal DIRS-based FPJ, where the transmit power per LU is -2 dBm (high transmit power).}
    \label{ResFig41}
\end{figure}

\begin{figure}[!t]
    \centering
    \subfloat{
        \includegraphics[scale=0.4]{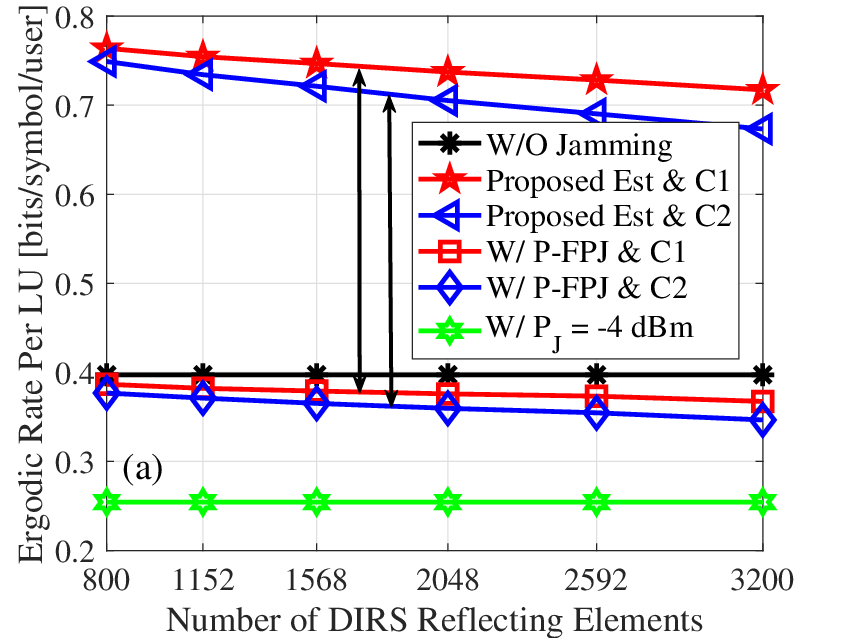}}\\
    \subfloat{
        \includegraphics[scale=0.4]{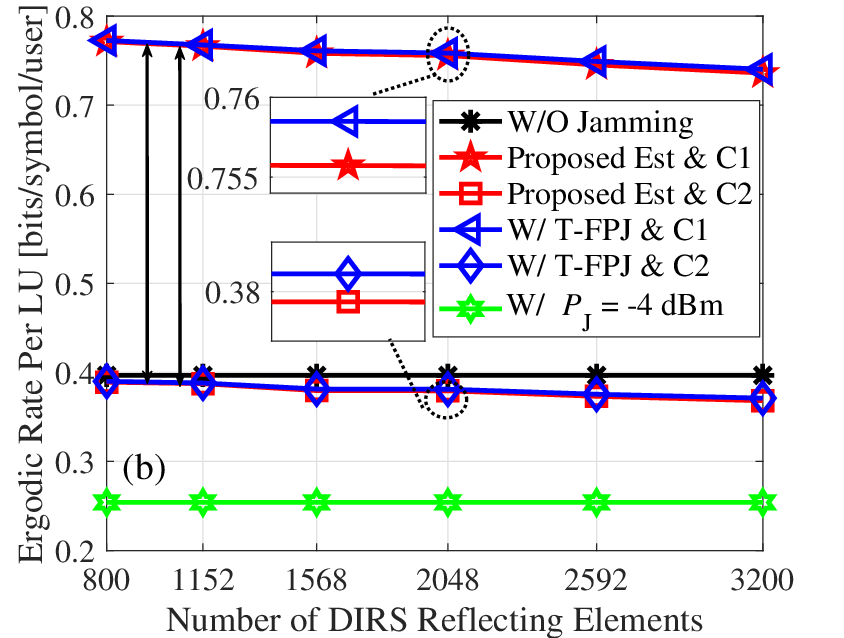}}  
    \caption{Relationship between the ergodic rate per LU and the number of DIRS reflecting elements for different benchmarks under jamming attacks launched by (a) the persistent DIRS-based FPJ and (b) the temporal DIRS-based FPJ, where the transmit power per LU is -14 dBm (low transmit power).}
    \label{ResFig42}
\end{figure}

Similarly, Fig.~\ref{ResFig3l} shows the relationship between the rates per LU and the amount of feedback at -14 dBm transmit power. As mentioned above, the rates resulting from the anti-jamming precoder are even better than the rates from an MU-MISO system without jamming attacks in the low transmit power domain, approximately twice as high. Based on Fig.~\ref{ResFig2h} and Fig.~\ref{ResFig3l}, the rates resulting from the proposed anti-jamming precoder are obtained by feeding back the received power only twice.

\subsubsection{Ergodic LU Rate Versus Number of DIRS Reflecting Elements Based on Estimated Statistical Characteristics}
Figs.~\ref{ResFig41} and~\ref{ResFig42} show the influence of the number of DIRS reflecting elements at high (-2 dBm) and low (-14 dBm) transmit power, respectively. Based on Propositions~\ref{Proposition1} and~\ref{Proposition2}, the variances of the DIRS-based ACA and hence the jamming impact for both the persistent DIRS-based FPJ and the temporal DIRS-based FPJ become significant as the number of DIRS reflecting elements increases. 
However, the proposed anti-jamming precoder always mitigates the jamming attacks launched by the persistent DIRS-based FPJ and the temporal DIRS-based FPJ, and the proposed anti-jamming precoder can even improve the rates in the low transmit power domain by exploiting the DIRS-jammed channels.
The DIRS-based ACA interference is a type of IUI, and thus it can be seen from Fig.~\ref{ResFig42} that neither the persistent DIRS-based FPJ nor the temporal DIRS-based FPJ can effectively jam an MU-MISO system with low transmit power, even when the number of the DIRS reflecting elements is large.

\begin{figure}[!t]
    \centering
    \subfloat{
        \includegraphics[scale=0.4]{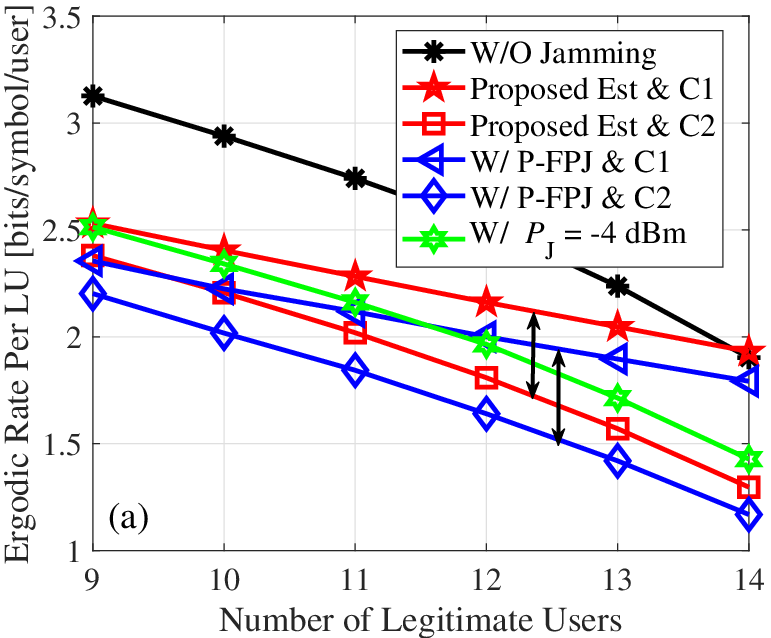}}\\
    \subfloat{
        \includegraphics[scale=0.4]{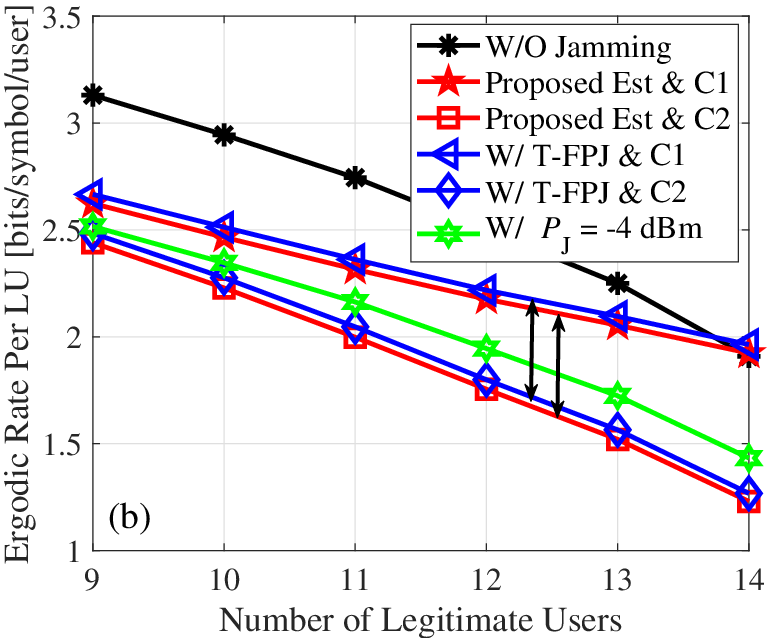}}  
    \caption{Relationship between the ergodic rate per LU and the number of LUs for different benchmarks under jamming attacks launched by (a) the persistent DIRS-based FPJ and (b) the temporal DIRS-based FPJ, where the transmit power per LU is -2 dBm (high transmit power).}
	\label{ResFig51}
\end{figure}

\begin{figure}[!t]
    \centering
    \subfloat{
        \includegraphics[scale=0.4]{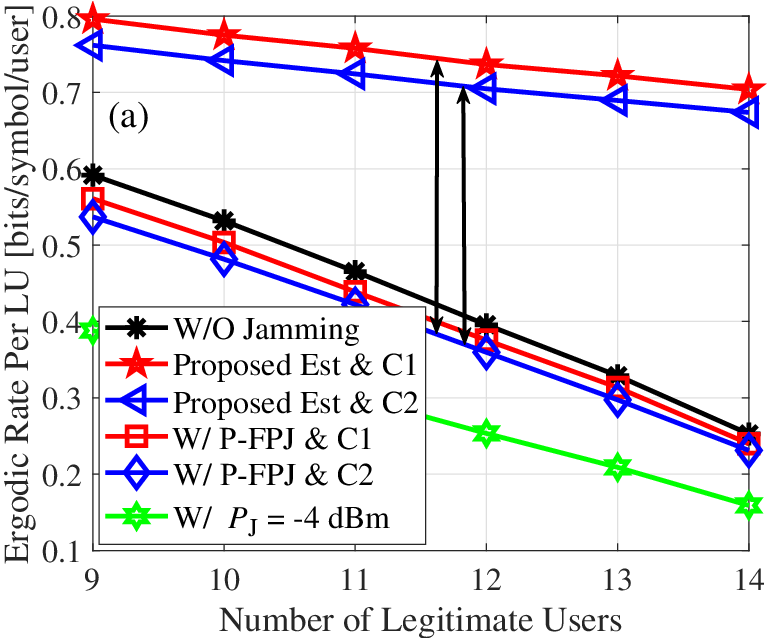}}\\
    \subfloat{
        \includegraphics[scale=0.4]{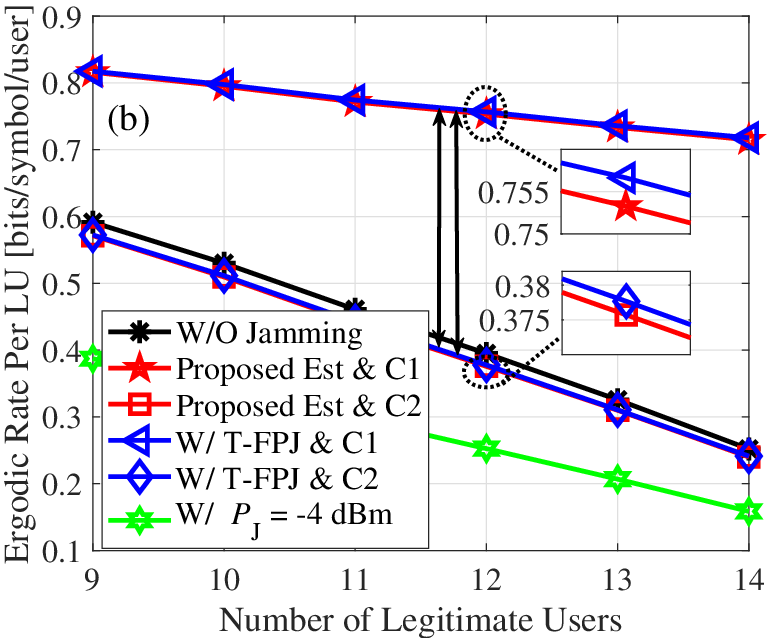}} 
    \caption{Relationship between the ergodic rate per LU and the number of LUs for different benchmarks under jamming attacks launched by (a) the persistent DIRS-based FPJ and (b) the temporal DIRS-based FPJ, where the transmit power per LU is -14 dBm (low transmit power).}
	\label{ResFig52}
\end{figure}

\begin{figure}[!t]
    \centering
    \subfloat{
        \includegraphics[scale=0.4]{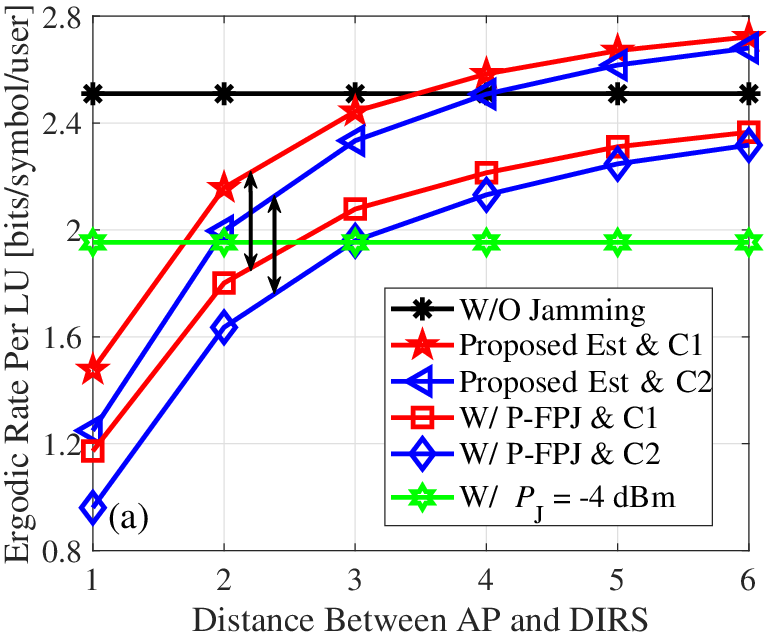}}\\
    \subfloat{
        \includegraphics[scale=0.4]{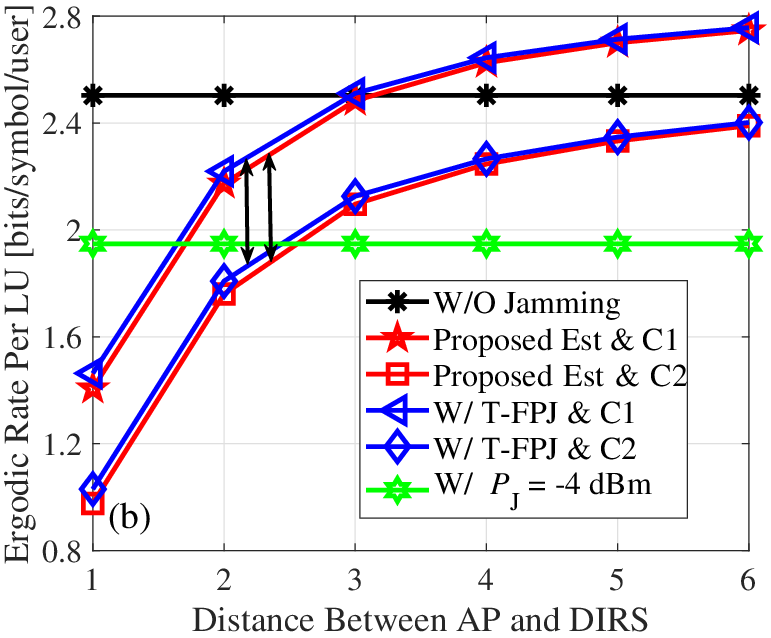}} 
    \caption{Relationship between the ergodic rate per LU and the AP-DIRS distance for different benchmarks under jamming attacks launched by (a) the persistent DIRS-based FPJ and (b) the temporal DIRS-based FPJ, where the transmit power per LU is -2 dBm (high transmit power).}
	\label{ResFig61}
\end{figure}

\begin{figure}[!t]
    \centering
    \subfloat{
        \includegraphics[scale=0.4]{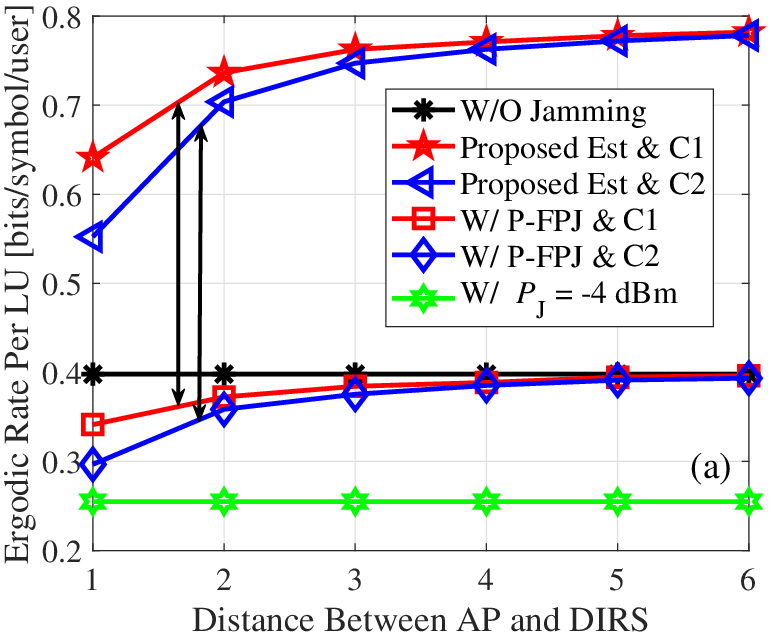}}\\
    \subfloat{
        \includegraphics[scale=0.4]{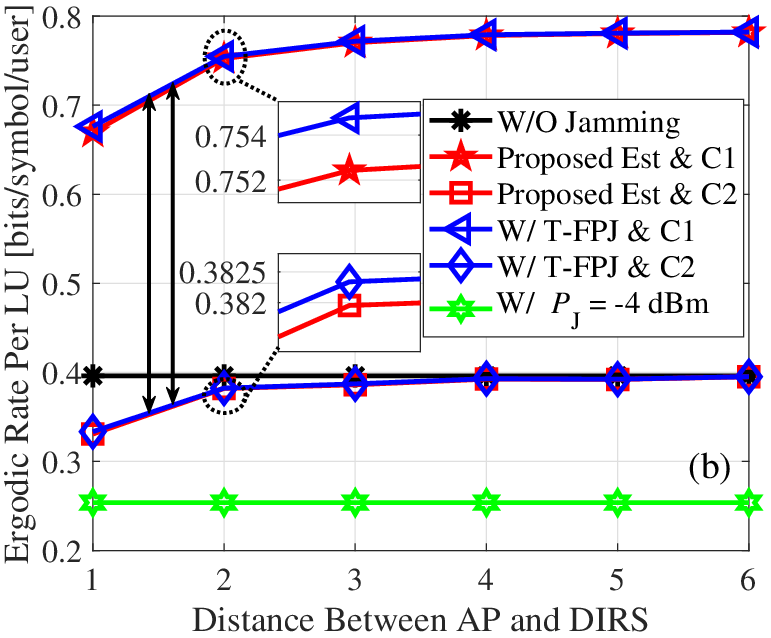}} 
    \caption{Relationship between the ergodic rate per LU and the AP-DIRS distance for different benchmarks under jamming attacks launched by (a) the persistent DIRS-based FPJ and (b) the temporal DIRS-based FPJ, where the transmit power per LU is -14 dBm (low transmit power).}
	\label{ResFig62}
\end{figure}

\subsubsection{Ergodic LU Rate Versus Number of Legitimate Users Based on Estimated Statistical Characteristics}
Figs.~\ref{ResFig51} and~\ref{ResFig52} show the ergodic rate per LU versus the number of LUs at high (-2 dBm) and low (-14 dBm) transmit power, respectively. The rates resulting from all benchmarks decrease with the number of LUs due to the increase in IUI and the decrease in available MIMO gain. However, as is illustrated in the figures, a unique property of the persistent DIRS-based FPJ and the temporal DIRS-based FPJ is that their jamming impact does not decrease as the number of LUs increases, but actually becomes more severe.
Fortunately, the mitigation generated by the anti-jamming precoder becomes more effective as the number of LUs increases.
Consequently, the gain generated from the jammed channel becomes more significant due to the anti-jamming precoder.
In addition, the difference between the rate achieved without any jamming and the rate obtained with active jamming attacks gradually decreases as the number of LUs increases. This is due to the fact that the increase in IUI detracts from the rates, while at the same time weakening the impact of AJ.

\subsubsection{Ergodic LU Rate Versus Distance Between Legitimate AP and DIRS Based on Estimated Statistical Characteristics}
In Figs.~\ref{ResFig61} and~\ref{ResFig62}, the impact of the DIRS location on the ergodic rates is illustrated at high (-2 dBm) and low (-14 dBm) transmit power, respectively. The greater the AP-DRIS distance, the greater the large-scale
channel fading ${\mathscr{L}}_{\rm G}$ in the AP-DIRS channel. According to Propositions~\ref{Proposition1} and~\ref{Proposition2}, the jamming impacts of the persistent DIRS-based FPJ and the temporal DIRS-based FPJ are weakened due to increased AP-DIRS distance $d_{\rm{AD}}$. From Fig.~\ref{ResFig61}, it is seen that the proposed anti-jamming precoder can achieve a rate similar to the case without jamming when the impact of the DIRS-based FPJs is weak due to the increased AP-DIRS distance $d_{\rm{AD}}$. In particular, for $d_{\rm{AD}} > 3$, the proposed anti-jamming precoder can completely compensate for the performance degradation imposed by the DIRS-based FPJs. As the distance $d_{\rm{AD}}$ continues to increase (the jamming impacts are weaker), the rates resulting from the proposed anti-jamming precoder are even better than those obtained without jamming. This is because that the gain obtained from the DIRS-based channels is greater than the degradation due to the jamming attacks.

\section{Conclusions}\label{Conclu}
In this paper, we have addressed the significant threats posed by DIRS-based FPJs. To this end, a novel anti-jamming precoder was developed that can be implemented by exploiting only the statistical characteristics of the DIRS-jammed channels instead of their instantaneous CSI.
 Our theoretical analysis and numerical results lead to the following conclusions, which raise concerns about the significant physical risks posed by DIRS-based FPJs.
\begin{enumerate}
\item DIRS-based FPJs launch jamming attacks by introducing multi-user ACA interference generated by the DIRS, and thus they can jam LUs with neither jamming power nor knowledge of the LU CSI. Increasing the transmit power at the legitimate AP will not reduce the jamming impacts of the DIRS-based FPJs and will actually make them more deleterious. Furthermore, the DIRS-based FPJs can defeat existing anti-jamming techniques such as spread spectrum, frequency-hopping, and MIMO interference cancellation.
\item The elements of the DIRS-based ACA channels follow a complex Gaussian distribution with zero mean and variance ${{{\mathscr{L}}\!_{{\rm G}}}{{\mathscr{L}}\!_{{\rm I},k}}{N\!_{\rm D}} {\overline \alpha} }$, and their variance is related to the distribution of the random DIRS phase shifts, since the gain value of each DIRS reflecting element is a function of its corresponding phase shift. The jamming impact of the persistent DIRS-based FPJ is more severe than that of the temporal DIRS-based FPJ, and in addition, the persistent DIRS-based FPJ does not require synchronization. Therefore, the persistent DIRS-based FPJ is more harmful than the temporal DIRS-based FPJ for an MU-MISO system.
\item Based on the derived distribution of the ACA channel, an anti-jamming precoder is presented that can achieve the maximum SJNR. In particular, for an MU-MISO system operating with low transmit power, the proposed anti-jamming precoder causes both the persistent DIRS-based FPJ and the temporal DIRS-based FPJ to not only fail to jam the LUs, it actually improves the SJNRs of the LUs due to the additional channel paths they provide. To obtain the statistical characteristics in practice, a data frame structure is then designed for the legitimate AP to estimate the statistical characteristics, which only requires the LUs to feed back their received power once or twice to the legitimate AP.
\end{enumerate}

\appendices
\section{Proof of Proposition~\ref{Proposition1}}\label{AppendixA}
According to~\eqref{ChannelACA}, the DIRS-based ACA channel ${\bf{H}}_{\rm {\!A\!C\!A}}$ can be written as ${{\bf{H}}_{\rm{\!A\!C\!A}}} =  {{\bf{H}}_{{\rm{I}}} \!\left( {{\bf \Phi}_{\!D\!T}} - {{\bf \Phi}_{\!R\!P\!T}} \right)\!{{\bf{G}}}}$. Consequently, the elements $\left[{\bf{H}}_{\rm {\!A\!C\!A}}\right]\!_{k,n}$ are given by
\begin{alignat}{1}
    \nonumber
    &{\left[ {{{\bf{H}}_{{\rm{ACA}}}}} \right]_{k,n}} = \\ \nonumber
    & \sqrt {\frac{{{\varepsilon _n}{{{\mathscr{L}}}_{\rm{G}}}{{\mathscr{L}}_{{\rm{I}},k}}}}{{{\varepsilon _n} + 1}}}{\widehat {{\boldsymbol{h}}}_{{\rm I},k}} \!\odot\! \left({\boldsymbol{\varphi}}(t_{\!D\!T}) - {\boldsymbol{\varphi}}(t_{\!R\!R\!T})\right)\left[{\widehat {\bf{G}}}^{{\rm{LOS}}}\right]_{:,n}  \\
    &+ \sqrt {\frac{{{{{\mathscr{L}}}_{\rm{G}}}{{\mathscr{L}}_{{\rm{I}},k}}}}{{{\varepsilon _n} + 1}}}{\widehat {{\boldsymbol{h}}}_{{\rm I},k}} \!\odot\! \left({\boldsymbol{\varphi}}(t_{\!D\!T}) - {\boldsymbol{\varphi}}(t_{\!R\!R\!T})\right)\left[{\widehat {\bf{G}}}^{{\rm{NLOS}}}\right]_{:,n},
    \label{RewriHACAele}
\end{alignat}
where $k=1,\cdots,K$, $n=1,\cdots,N_{\rm A}$, and $\odot$ represents the Hadamard product.

Furthermore, $\left[{\bf{H}}_{\rm {\!A\!C\!A}}\right]\!_{k,n}$ can be reduced to
\begin{alignat}{1}
    \nonumber
    &\!\!\!\!\!\!\!\!\!\!\!\!\!{\left[ {{{\bf{H}}_{{\rm{ACA}}}}} \right]_{k,n}} = \\\nonumber
    &\sqrt {\frac{{{\varepsilon _n}{{{\mathscr{L}}}_{\rm{G}}}{{\mathscr{L}}_{{\rm{I}},k}}}}{{{\varepsilon _n} + 1}}} \sum\limits_{r = 1}^{{N_{\rm{D}}}} \!\Big(\!\left[{\widehat {{\boldsymbol{h}}}_{{\rm I},k}}\right]_{r} \!\!\!\big( {\alpha _r}({t_{DT}}){e^{j{\varphi _r}({t_{DT}})}} \\\nonumber
    &- {\alpha _r}({t_{RPT}}){e^{j{\varphi _r}({t_{RPT}})}} \big) \!\!\left[{\widehat {\bf{G}}}^{{\rm{LOS}}}\right]_{r,n}\!\Big) \\ \nonumber
    &+\sqrt {\frac{{{{{\mathscr{L}}}_{\rm{G}}}{{\mathscr{L}}_{{\rm{I}},k}}}}{{{\varepsilon _n} + 1}}}\sum\limits_{r = 1}^{{N_{\rm{D}}}} \!\Big(\!\left[{\widehat {{\boldsymbol{h}}}_{{\rm I},k}}\right]_{r} \!\!\!\big( {\alpha _r}({t_{DT}}){e^{j{\varphi _r}({t_{DT}})}} \\
    &- {\alpha _r}({t_{RPT}}){e^{j{\varphi _r}({t_{RPT}})}} \big) \!\!\left[{\widehat {\bf{G}}}^{{\rm{NLOS}}}\right]_{r,n}\!\Big)
    \label{RewriHACAele1} \\
    &=  {\sqrt{\frac{{{\varepsilon _n}{{{\mathscr{L}}}_{\rm{G}}}{{\mathscr{L}}_{{\rm{I}},k}}}}{{{\varepsilon _n} + 1}}}} {\sum\limits_{r = 1}^{{N_{\rm{D}}}} {{a_r}}} + {\sqrt {\frac{{{{{\mathscr{L}}}_{\rm{G}}}{{\mathscr{L}}_{{\rm{I}},k}}}}{{{\varepsilon _n} + 1}}}}{\sum\limits_{r = 1}^{{N_{\rm{D}}}} {{b_r}}}.
    \label{RewriHACAele2}
\end{alignat}
Conditioned on the fact that the random variables in~\eqref{RewriHACAele1} are independent, we have ${\mathbb{E}}\!\left[ a_r \right] = {\mathbb{E}}\!\left[ b_r \right] = 0$.
Furthermore, the variance of $a_r$ is
\begin{equation}
    \begin{split}
        {\rm{Var}}\!\left[ {a_r} \right] = {\mathbb{E}}\!\left[ {a_r} {a_r^H}  \right] =  {\mathbb{E}} \Big[   {{{\left| {{\alpha _r}\!({t_{\!D\!T}})} \right|}^2} + {{\left| {{\alpha _r}\!({t_{\!R\!P\!T}})} \right|}^2}}  \\
        - {\alpha _r}\!({t_{\!D\!T}}){\alpha _r}\!({t_{\!R\!P\!T}})\cos \!\left( {{\varphi _r}\!({t_{\!D\!T}}) - {\varphi _r}\!({t_{\!R\!P\!T}})} \right) \Big].
    \end{split}
    \label{Varb}
\end{equation}
Based on the definition in Section~\ref{DIRSFPJ}, the variance expressed in~\eqref{Varb} reduces to
\begin{equation}
    {\rm{Var}}\!\left[ {a_r} \right]  \!=\!  \sum\limits_{i1 = 1}^{{2^b}} {\sum\limits_{i2 = 1}^{{2^b}} \!{ {P_{i1}}{P_{i2}} \!\left( {\mu _{i1}^2} \!+\! {\mu _{i2}^2} \!-\! 2{\mu _{i1}}{\mu _{i2}}\cos ( {{\theta _{i1}} \!-\! {\theta _{i2}}}  ) \right)  } },
\label{VaraVau}
\end{equation}
where ${\mu _{i1}}$, ${\mu _{i2}} \in \Omega$, and ${\theta _{i1}}$, ${\theta _{i2}} \in \Theta$. Similarly, the variance of $b_r$ can be derived as
\begin{equation}
    {\rm{Var}}\!\left[ {b_r} \right]  \!=\!  \sum\limits_{i1 = 1}^{{2^b}} {\sum\limits_{i2 = 1}^{{2^b}} \!{ {P_{i1}}{P_{i2}} \!\left( {\mu _{i1}^2} \!+\! {\mu _{i2}^2} \!-\! 2{\mu _{i1}}{\mu _{i2}}\cos ( {{\theta _{i1}} \!-\! {\theta _{i2}}}  ) \right)  } }.
    \label{VarbVau}
\end{equation}

Based on the Lindeberg-L$\acute{e}$vy central limit theorem, we have
\begin{alignat}{1}
\frac{{\sum\limits_{r = 1}^{N_{\rm D}} {{a_r}} }}{{\sqrt {N_{\rm D}} }} & \mathop  \to \limits^{\rm{d}} \mathcal{CN}\left( {0,{\overline \alpha}} \right),\;{\rm{as}}\; {N_{\rm D}} \to \infty,
\label{aCLTeq}\\
\frac{{\sum\limits_{r = 1}^{N_{\rm D}} {{b_r}} }}{{\sqrt {N_{\rm D}} }} & \mathop  \to \limits^{\rm{d}} \mathcal{CN}\left( {0,{\overline \alpha}} \right),\;{\rm{as}}\; {N_{\rm D}} \to \infty,
\label{bCLTeq}
\end{alignat}
where 
    $\overline \alpha = \sum\nolimits_{i1 = 1}^{{2^b}} \sum\nolimits_{i2 = 1}^{{2^b}}  {P_{i1}}{P_{i2}} \big( {\mu _{i1}^2} + {\mu _{i2}^2} - 2{\mu _{i1}}{\mu _{i2}}$ $\cos ( {{\theta _{i1}} - {\theta _{i2}}} ) \big)$.

Consequently, the elements $\left[{\bf{H}}_{\rm {\!A\!C\!A}}\right]\!_{k,n}$ in~\eqref{RewriHACAele2} follow
\begin{equation}
{\left[ {{\bf H}_{\rm {\!A\!C\!A}}} \right]_{k,n}} \mathop  \to \limits^{\rm{d}} \mathcal{CN}\!\left( {0,  {{{\mathscr{L}}\!_{{\rm G}}}{{\mathscr{L}}\!_{{\rm I},k}}{N\!_{\rm D}}{\overline \alpha} } } \right), \forall k,n.
\label{HDStare}
\end{equation}
\section{Proof of Proposition~\ref{Proposition2}}\label{AppendixB}
If the DIRS only changes its reflection coefficients during the \emph{DT} phase and remains silent during the \emph{RPT} phase, we have ${\alpha _r}({t_{\!R\!P\!T}}) = 0$ and ${\varphi _r}(t_{\!D\!T}) \in {\cal R}\!\left(\Theta\right)$, where $r=1,\cdots, N_{\rm D}$. Therefore, the overall combined channels in the $RPT$ and $DT$ phases are reduced to ${\bf H}_{\!R\!P\!T} = {\bf 0}$ and ${\bf H}_{\!D\!T} = {\bf H}_{\rm I}{{{\bf \Phi}}_{\!R\!P\!T}}{\bf G}$, respectively.
Consequently, the elements of the DIRS-based ACA channel ${\bf{H}}_{\rm {\!A\!C\!A}}$ in~\eqref{ChannelACA} are reduced to
\begin{alignat}{1}
    \nonumber
    {\left[ {{{\bf{H}}_{{\rm{ACA}}}}} \right]_{k,n}} = &\sqrt {\frac{{{\varepsilon _n}{{{\mathscr{L}}}_{\rm{G}}}{{\mathscr{L}}_{{\rm{I}},k}}}}{{{\varepsilon _n} + 1}}}{\widehat {{\boldsymbol{h}}}_{{\rm I},k}} \!\odot\! {\boldsymbol{\varphi}}(t_{\!D\!T})\!\left[{\widehat {\bf{G}}}^{{\rm{LOS}}}\right]_{:,n}  \\
    &+ \sqrt {\frac{{{{{\mathscr{L}}}_{\rm{G}}}{{\mathscr{L}}_{{\rm{I}},k}}}}{{{\varepsilon _n} + 1}}}{\widehat {{\boldsymbol{h}}}_{{\rm I},k}} \!\odot\! {\boldsymbol{\varphi}}(t_{\!D\!T})\!\left[{\widehat {\bf{G}}}^{{\rm{NLOS}}}\right]_{:,n}.
    \label{RewriHACAele18}
\end{alignat}

Similar to~\eqref{RewriHACAele1} and~\eqref{RewriHACAele2}, we rewrite~\eqref{RewriHACAele18} as
\begin{alignat}{1}
    \nonumber
    &{\left[ {{{\bf{H}}_{{\rm{ACA}}}}} \right]_{k,n}} = \\ \nonumber
    & \sqrt {\frac{{{\varepsilon _n}{{{\mathscr{L}}}_{\rm{G}}}{{\mathscr{L}}_{{\rm{I}},k}}}}{{{\varepsilon _n} + 1}}} \sum\limits_{r = 1}^{{N_{\rm{D}}}} \!\left({\!\left[{\widehat {{\boldsymbol{h}}}_{{\rm I},k}}\right]_{r} \!\!{\alpha _r}({t_{DT}}){e^{j{\varphi _r}({t_{DT}})}}\!\!\left[{\widehat {\bf{G}}}^{{\rm{LOS}}}\right]_{r,n}}\!\right) \\
    & +\sqrt {\frac{{{{{\mathscr{L}}}_{\rm{G}}}{{\mathscr{L}}_{{\rm{I}},k}}}}{{{\varepsilon _n} + 1}}}\sum\limits_{r = 1}^{{N_{\rm{D}}}} \!\left({\!\left[{\widehat {{\boldsymbol{h}}}_{{\rm I},k}}\right]_{r} \!\!{\alpha _r}({t_{DT}}){e^{j{\varphi _r}({t_{DT}})}} \!\!\left[{\widehat {\bf{G}}}^{{\rm{NLOS}}}\right]_{r,n}}\!\right)
    \label{Rewri18HACAele1} \\
    &=  {\sqrt{\frac{{{\varepsilon _n}{{{\mathscr{L}}}_{\rm{G}}}{{\mathscr{L}}_{{\rm{I}},k}}}}{{{\varepsilon _n} + 1}}}} {\sum\limits_{r = 1}^{{N_{\rm{D}}}} {{c_r}}} + {\sqrt {\frac{{{{{\mathscr{L}}}_{\rm{G}}}{{\mathscr{L}}_{{\rm{I}},k}}}}{{{\varepsilon _n} + 1}}}}{\sum\limits_{r = 1}^{{N_{\rm{D}}}} {{d_r}}}.
    \label{Rewri18HACAele2}
\end{alignat}
It is easy to show that the expectations of $c_r$ and $d_r$ are equal to zero. Furthermore, their variances are given by
\begin{equation}
{\rm{Var}}\!\left[c_r\right] = {\rm{Var}}\!\left[d_r\right] = {\mathbb{E}}\!\left[ {{{\left| {{\alpha _r}({t_{DT}})} \right|}^2}} \right] =\sum\limits_{i = 1}^{{2^b}} {{P_i}} \mu _i^2.
\label{VarHACAele18}
\end{equation}

Based on the Lindeberg-L$\acute{e}$vy central limit theorem, the elements $\left[{\bf{H}}_{\rm {\!A\!C\!A}}\right]\!_{k,n}$ in~\eqref{Rewri18HACAele2} satisfy
\begin{equation}
{\left[ {{\bf H}_{\rm {\!A\!C\!A}}} \right]_{k,n}} \mathop  \to \limits^{\rm{d}} \mathcal{CN}\!\left( {0,  {{{\mathscr{L}}\!_{{\rm G}}}{{\mathscr{L}}\!_{{\rm I},k}}{N\!_{\rm D}}{\overline \alpha} } } \right), \forall k,n,
\label{HDStare18}
\end{equation}
where $\overline \alpha = {{\sum\nolimits_{i = 1}^{{2^b}} {P_i}{\mu_i^2} }}$.

\end{document}